\documentclass{aa}

\usepackage{graphicx}
\usepackage{txfonts}
\usepackage{xspace}
\usepackage{mathrsfs}
\usepackage{comment}
\usepackage{lineno}
\usepackage{natbib,twoopt}
\usepackage{multicol}
\usepackage[breaklinks=true]{hyperref}

\newcommand\fermi{\textit{Fermi}-LAT\xspace}

\newcommand{\beq}{\begin{equation}}
\newcommand{\eeq}{\end{equation}}
\newcommand{\der}{\mathrm{d}}
\newcommand{\dgr}{\ensuremath{^\circ}\,}
\newcommand{\hii}{H{\tiny II}}
\newcommand{\nHii}{\ensuremath{n_{HII}}\,}
\newcommand{\RHii}{\ensuremath{R_{HII}}\,}

\def\cherenkov{Cherenkov}

\def\gr{$\gamma$-ray}
\def\grs{$\gamma$-rays}

\begin{document}

\title{Embedded star clusters as sources of high-energy cosmic rays }
\subtitle{Modelling and constraints}
\titlerunning{Embedded star clusters, sources of cosmic rays?}

\author{G. Maurin\inst{1}, A. Marcowith\inst{2}, N. Komin\inst{3}, F. Krayzel\inst{1},
          \and G. Lamanna\inst{1} }

\institute{\inst{1}Laboratoire d’Annecy-le-Vieux de Physique des Particules, Université Savoie Mont-Blanc, CNRS/IN2P3, F-74941 Annecy-le-Vieux, France\\
                  \inst{2}Laboratoire Univers et Particules de Montpellier, Université Montpellier 2, CNRS/IN2P3, CC 72, Place Eugène Bataillon, F-34095
Montpellier Cedex 5, France\\
             \inst{3}School of Physics, University of the Witwatersrand, 1 Jan Smuts Avenue, Braamfontein, Johannesburg, 2050 South Africa\\
             }

\offprints{G.~Maurin \\
     \email{\href{mailto: gilles.maurin@lapp.in2p3.fr}{gilles.maurin@lapp.in2p3.fr}}
     }

   \date{Received DAY MONTH 2016; accepted DAY MONTH 2016}

\abstract
   {Massive stars are mainly found in stellar associations. These massive star clusters occur in the heart of giant molecular clouds. The strong stellar wind activity in these objects generates large bubbles and induces collective effects that could accelerate particles up to high energy and produce \grs. The best way to input an acceleration origin to the stellar wind interaction in massive stellar cluster is to observe young massive star clusters in which no supernova explosion has occurred yet.}
   {This work aims to constrain the part of stellar wind mechanical energy that is converted into energetic particles using the sensitivity of the ongoing \fermi\ instrument. This work further provides detailed predictions of expected \gr\ fluxes in the view of the on-set of the next generation of imaging atmospheric Cherenkov telescopes.}
   {A one-zone model where energetic particles are accelerated by repeated interactions with strong supersonic shocks occurring in massive star clusters was developed.  The particle escape from the star cluster and subsequent interaction with the surrounding dense material and magnetic fields of the \hii\ region was computed. We applied this model to a selection of eight embedded star clusters constricted by existing observations. We evaluated the \gr\ signal from each object, combining both leptonic and hadronic contributions. We searched for these emissions  in the \fermi\ observations in the energy range from 3 to 300\,GeV and compared them to the sensitivity of the Cherenkov Telescope Array (CTA).}
   {No significant \gr\ emission from these star clusters has been found. Less than 10\% of stellar wind luminosities are supplied to the relativistic particles. Some clusters even show acceleration efficiency of less than 1\%. The CTA would be able to detect \gr\  emission from several clusters in the case of an acceleration efficiency of close to one percent.}
   {}

   \keywords{Star forming region, gamma ray astronomy, Cosmic Rays, Embedded Star Clusters, Fermi-LAT, CTA }

\maketitle

\section{Introduction}
The shells of supernova remnants (SNRs) are considered to be the main acceleration sites of galactic cosmic rays \citep[GCRs; see e.g.][]{Drury01}. The detection of radio synchrotron radiation from SNR shells has already proved that electrons with relativistic energies are accelerated in these objects. Recent X-ray synchrotron emission observed in several historical SNRs have been shown to result from the acceleration of electrons up to TeV energies in a magnetic field amplified by two orders of magnitude with respect to standard interstellar values \citep{Parizot06}. The magnetic field amplification is likely caused by the acceleration of protons and heavier ions \citep{Bell01}. Even if a definite proof of a leptonic or hadronic origin of the \gr\ emission does not exist yet, the presence of TeV photons detected from these shells \citep[for a review of TeV-detected SNRs see e.g.][]{Blois} further supports the hypothesis that SNRs are sources of GCRs. This emission can be interpreted either as inverse Compton  (IC) upscattering of low-energy photons by the relativistic electrons or as a product of decaying pions produced by hadronic interactions of CRs with the ambient shocked material.

Other potential sources of energetic particles (to become CRs) exist, however, and at this stage still cannot be excluded \citep[see again][]{Drury01}. Similarly to SNRs, CRs can also be accelerated in the shells of superbubbles. The superbubbles are blown into the interstellar medium by the combined effects of stellar winds produced by associations of massive stars and successive supernova explosions. So far, GeV and TeV \gr\ emissions have been detected in a few superbubbles; for instance, the \fermi\ instrument has detected GeV radiation from Cygnus~OB2 in the Milky Way \citep{ackermann2011cocoon} and the H.E.S.S. \cherenkov\ telescope has detected a signal from  30~Dor~C in the Large Magellanic Cloud \citep{Abramowski:2015rca} and from Westerlund~1 \citep{Abramowski12} in the Milky Way. This \gr\ emission is evidence that particle acceleration is ongoing in these objects. Although in the cases cited above several supernova explosions have likely occurred and it is possible that embedded SNRs are responsible for (or contribute to) particle acceleration in these objects \citep{Bykov13, Ferrand10, Butt10, Ferrand08}.

Most of the massive stars are born, live, and die in clusters. Rich and dense populations of massive stars can in fact be found in massive star clusters (MSC), which host strong supersonic winds with velocities $\sim$ 1000 km/s or even beyond. These winds can interact each other directly \citep{Parizot04}. The mechanical energy imparted in the winds of the most massive stars ($\geq$ O4-type) over their lifetime is equivalent to the mechanical energy deposited by a supernova explosion. Hence, if as stated above, SNRs are among the possible sites of cosmic particle acceleration; direct interactions of massive stellar winds in a cluster can also drive particle acceleration \citep{Bykov01}. In that view, the recent detection of \gr\ emission from Eta Carina by the Fermi collaboration \citep{Abdo10} modulated by its orbital period \citep{Reitberger15} suggests that wind-wind interaction in these systems can accelerate particles up to high energies. In this work it is assumed, following the model of \citet{Klepach00}, that energetic particles are accelerated in MSC owing to collective acceleration produced by multiple shock encounters.
The contribution of massive stars to the production of energetic particles is tested. To that aim, a simple one-zone model is developed, in which \gr\ radiations are produced by relativistic particles accelerated in a MSC interacting with the matter concentrated in the dense surrounding \hii\ region. This model is different from the approach of \citet{Domingo06} and \citet{Torres04}, where \gr\ radiations are produced  in the MSC. From different catalogues a list of young MSCs, where no supernova has occurred yet, was selected. As a result of their high density, the radiation produced by the \hii\ regions is significantly higher and much better constrained than in MSCs. One should keep in mind that the one-zone model considered here is conservative as each MSC could also contribute to the global \gr\ emission. This model is similar and complementary to the work of \citet{Bednarek07}. For different sets of sources, constraints on the conversion parameter of the wind power into energetic particles using the \fermi\ observations in the energy range from 3 to 300\,GeV are obtained. This work also provides detailed predictions of expected \gr\ fluxes for the future Cherenkov Telescope Array (CTA). The reliability of the results is finally discussed with respect to CR transport models in \hii\ regions. The \gr\ emission from young star clusters and stellar-wind bubbles (SWBs) would be the best evidence for particle acceleration originated by stellar wind interactions.
The article is organized as follows. In section 2 the model of particle acceleration in embedded star clusters is presented. Section 3 details the electron and proton spectra as well as the \gr\ emission obtained from the model. The selection of the MSC sample, the physical parameters of the MSCs, and the search for \gr\ emission are discussed in section 4. Section 5 discusses different tests of the model with respect to the MSC sample data and section 6 is a discussion of the efficiency of particle acceleration in young MSCs. The conclusions and some perspectives are outlined in section 7.

\section{Particle acceleration model}
\label{modelexplain}
\begin{figure}[ht]
\begin{center}
\includegraphics[width=5cm]{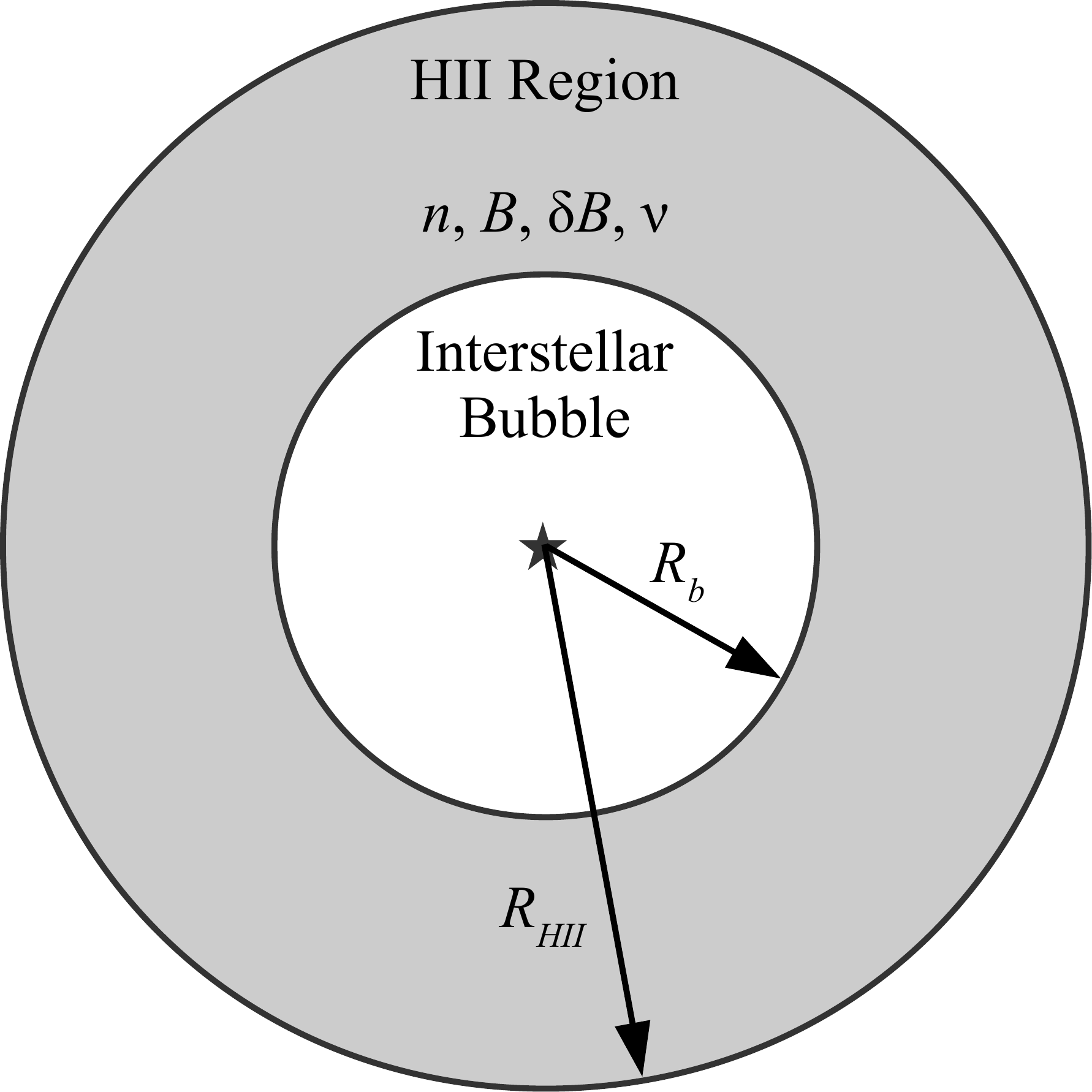}
\caption{Spherical model of a star cluster embedded in a \hii\ region. The Interstellar Bubble (or region 1) as a radius $R_b$. The \hii\ region (or region 2) has an outer radius of \RHii\ and is characterized by the density $n$,  magnetic field strength $B$, intensity of the magnetic turbulence level $\delta B,$ and index of turbulence $\nu$. The Interstellar Bubble is much smaller than the \hii\ region, $R_b << \RHii$.
}
\label{fig:geomod}
\end{center}
\end{figure}

The main hypothesis of this model is to assume that a fraction $\xi$ of the total stellar wind luminosity $L_{w}$ released by the central cluster is converted into energetic particle power $P_p$. Namely, for protons and electrons, respectively, a specific fraction is introduced:
\begin{itemize}
\item[$\bullet$] For protons : $P_{p} = \xi_p \; L_{w}$
\item[$\bullet$] For electrons : $P_{e} = \xi_e L_{w} = K_{ep} \; \xi_p \; L_{w}$, where $K_{ep}$ is the electron-to-proton ratio.
\end{itemize}
The $L_w$ is assumed to be constant as function of the time and is the sum of the contribution of massive stars contained in the cluster.
The model has two main regions, which are shown in Fig.~\ref{fig:geomod}. The Interstellar Bubble is the central region containing the massive star cluster where particles are accelerated by collective shock acceleration, and the \hii\ region is the second region engulfing the stellar cluster that contains most of the mass of the system and hence the target material for energetic particles escaping the central cluster. For simplicity, a spherical geometry is assumed in this model.

The Interstellar Bubble corresponds to the region of the stellar wind in \citet{Weaver}; it should account for both unshocked and shocked stellar wind regions. The \hii\ region includes the shocked interstellar gas and the stellar-ionized interstellar gas \citep{Freyer03}. The size of the Interstellar Bubble, $R_b$, is identified with the stellar wind termination shock radius in \citet{Weaver} because it is difficult to derive the shocked stellar region observationally. The \RHii\ corresponds to the radius of the ionizing front of the \hii\ region associated with the cluster, and it is dependent of the cluster age \citep[see an estimation given by Eq.1 in][]{Freyer03}. Owing to the radiative luminosity of the cluster, the size of the bubble is negligible compared to the size of the \hii\ region although this is not always the case in reality. This implies that the \hii\ region is the main emission region.

As detailed by \citet{Aharonian}, a one-zone approximation to describe the particle spectrum in the \hii\ region is used,
\begin{equation}
\frac{\partial N}{\partial t} = \frac{\partial}{\partial E}\left(P(E) N\right) - \frac{N}{\tau_{esc}} + Q \ ,
\label{leaky}
\end{equation}
where $N$ is the energy density of the particles, $P(E)$ is the volume-averaged energy loss rate of the particles, $Q \equiv  Q(E,t)$ is the source term due to the acceleration process occurring in the Interstellar Bubble and $\tau_{esc} \equiv \tau_{esc}(E)$ is the escape time, which is the characteristic time needed for energetic particles to escape from the \hii\ region.

\subsection{Particle injection from the Interstellar Bubble}
The source term $Q(E,t)$ is linked with the energetic particle power by
\beq
P_{p} = \xi L_w = \int_V \int_{E_{inj}}^{E_{max}} Q(E,t) E \; \der E \; \der^{3} r.
\label{inject}
\eeq
Assuming a power-law injection spectrum, $Q(E) = Q_0 E^{-s}$, with an injection index typically chosen to be $s = 2$, this yields
\beq
\frac{\xi \; L_w}{V_{b}(t)} = \int_{E_{inj}}^{E_{max}} Q_0\left ( \frac{E}{E_{inj}} \right )^{-s} E \; \der E
\label{inject2}
,\eeq
where $E_{inj}$ is the injection kinetic energy of particles in the acceleration process and $E_{max}$ the maximum energy reached by particles by the collective acceleration process occurring in the cluster. The injection energy is fixed at $E_{inj}= (\sqrt{2} -1) m_p c^2$; it corresponds to an injection momentum of $m_p c$, where $m_p$ is the proton mass and c is the speed of light.

It is not the purpose of this simple approach to model the particle acceleration in the cluster., however, we briefly discuss the expected values of $E_{max}$ for protons and electrons below. Our estimations follow the wind models developed by \citet{Klepach00} (their model I and model II). The particles are accelerated by repeated shock acceleration associated with multiple interactions with wind termination shocks. If the shock filling factor is $f = 1$ then strong turbulence produced by shock-shock interaction controls the particle acceleration and transport \citep[see e.g.][]{Bykov01, Parizot04}. Multiple shock acceleration produces $E_{max}$ values that are higher than what is expected in isolated massive stars \citep[see e.g.][]{Voelk82}. In case the shock filling factor is $f < 1$ then the collective wind acceleration timescale is $t_{acc} = a R_w/(3 U_w f),$ where $R_w$ and $U_w$ are the termination shock radius and asymptotic velocity, respectively, and $a=3 r/(r-1)$ is controlled by the shock compression ratio $r$ (assumed to be 4 in this work). \citet{Klepach00} obtained maximum proton energies of $E_{max,p} \sim 2/\eta_D$  PeV, where $\eta_D=D_i/D_B$ is the ratio of the Bohm diffusion coefficient to the diffusion coefficient in the winds. We consider the possibility of having a diffusion that is close to Bohm diffusion and a value of $E_{max}$ in range 1-10 PeV is adopted.
If strong turbulence is generated in the MSC \citep{Bykov01}, maximum proton energies may reach 100 PeV.
The value 
$E_{max}$ is high enough to test possible \gr\ emission from massive star clusters. The electron maximum energy in the cluster is fixed by balancing the collective wind acceleration timescale with the synchrotron loss time. Accounting for uncertainties for $U_w$ and $R_w$ typical values of $E_{e,max}= 10-100$\,TeV are obtained for a  MSC magnetic field strength  $B_{MSC}=10-100 \ \rm{\mu G}$ \citep{Parizot04}. The maximum energy $E_{max}$ usually has only a weak influence on the results (see the discussion in section \ref{influence}).

Further on a homogeneous particle distribution is assumed, hence, the integral on the volume can be replaced by the volume of the bubble $V_{b}$. The bubble volume evolves with time. Following \citet{Weaver}, the size of the bubble increases as $R_b \propto t^{2/5}$. Hence $Q_0$ is also a function of the time and evolves as $Q_0 \propto t^{-6/5}$; i.e. more particles are injected at earlier times.

\subsection{Escape from the \hii\ region}
In the case of a three-dimensional (3D) isotropic diffusion process, the escape time can be written as
\begin{equation}
\label{tauesc}
\tau_{esc} (E, t) = \frac{\RHii(t)^{2}}{6 \; D(E)}
,\end{equation}
where \RHii\ is the \hii\ region radius, assumed to be constant in this work, and $D(E)$ the spatial diffusion coefficient inside the \hii\ region. The spatial diffusion coefficient is supposed to be proportional to the local galactic diffusion coefficient $D_0$. We fix the value at 3 GeV to $D_{0,3\rm{GeV}} \approx 4 \times 10^{28}\mathrm{cm}^{-1}\mathrm{s}^{-1}$, which is deduced from B/C ratio measurements in cosmic ray data \citep[see][]{Ptuskin06},
\begin{equation}
D(E) = \left ( \frac{D}{D_0} \right ) \times D_{0,3 \rm{GeV}} \times \left ( \frac{E}{3 \; \rm{GeV}} \right )^{2-\nu}
.\end{equation}
The coefficient is fixed by two parameters: the ratio $D/D_0$ and the index of turbulence $\nu$. We adopt a turbulence index of $\nu = 5/3$ in the case of a Kolmogorov turbulence or $\nu = 3/2$ for the Iroshnikov-Kraichnan turbulence. Following \citet{Marcowith}, the ratio $D/D_0$ can be derived from
\beq
D(E)= \frac{\nu }{ \pi (\nu-1)} \times \frac{\ell_c c }{ \eta} \times \left( \frac{r_l }{ \ell_c}\right)^{2-\nu}
\label{diff}
\eeq
to be
\begin{equation}
D/D_0 =
\left\{
  \begin{array}{l l}
\frac{D}{D_0}= \frac{1.2 \times 10^{-3}}{\eta} \times \left (\frac{\ell_{c,pc}}{B_{10}} \right )^{1/2} \quad \rm{if} \; \nu = 3/2 \\
\frac{D}{D_0}= \frac{1.2 \times 10^{-2}}{\eta} \times \ell_{c,pc}^{2/3} \; B_{10}^{-1/3} \quad \rm{if} \; \nu = 5/3
  \end{array} \right.
.\end{equation}
The following notations are used: $\eta= \delta B^2/(\delta B^2 + B_0^2) \in [0,1]$ is the turbulence level, $\delta B$ is the intensity of the magnetic turbulence level and $B_0$ the mean intensity of the magnetic field, $l_c$ the coherence length of the turbulent magnetic field, here fixed at 1 pc \citep[for details see][]{PhysRevD.90.041302}. Hereafter, two cases are studied: $\delta B = 0.1 B_0$, for low magnetic turbulence level, and $\delta B = B_0$, for a strongly turbulent case. $B_{10}$ is the mean magnetic field strength in units of $10 \mu \rm{G}$, $B_{10}=B_0/10 \mu \rm{G}$.

\subsection{Cooling in the \hii\ region}
As particles enter the \hii\ region, they suffer energy losses through several processes. For protons, the dominant loss process is pion production. Adiabatic losses do not play a role as the \hii\ region has an approximate stationary size with respect to the escape timescale of the energetic particles, and the Coulomb interaction and ionization losses are negligible at the energies under consideration. For electrons, the following loss processes are considered:\  \textit{i)} synchrotron losses that depend on the magnetic field in the \hii\ region; \textit{ii)} Bremsstrahlung losses that depend on the density of the \hii\ region; and \textit{iii)} inverse Compton losses using various photon fields: cosmic microwave background (CMB), Galactic photon field and the light from the cluster itself is added. This light is just a sum of black bodies from each O star of the stellar cluster. The content of O stars in the stellar cluster is known from the Galactic O-Star Catalog (see the section \ref{ClusterSelection} for the details of the cluster selection from this catalogue).

\subsection{Solution of the one-zone model}
According to \citet{Aharonian}, the energetic particle spectrum in the \hii\ region is given by the following analytical solution of the one-zone equation \ref{leaky}:
\begin{equation}
N(E,t) = \frac{1}{P(E)} \int_{0}^{t} P(E_t) \; Q(E_t,t') \; \exp \left( - \int_{t'}^{t} \frac{\der x}{\tau_{esc}(E_x)} \right) \der t'
\label{sol}
,\end{equation}
where $E_t$ is the energy of a particle at an instant $t' < t$ with an energy $E$ at an instant $t$ given by
\begin{equation}
t-t' = \int_{E_e}^{E_t} \frac{\der E'}{P(E')}.
\label{solbis}
\end{equation}

\subsection{Summary of the modelling procedure}
\label{paramresum}
The model depends on several parameters that are generally fixed or constrained by observations:\  \textit{i)} the age and distance from Earth are fixed by optical and/or infrared (IR) observations; \textit{ii)} the radius \RHii\ and density \nHii\ of the \hii\ region are fixed by optical observations and can be compared with the solutions outlined in \citet{Freyer03}; \textit{iii)} the stellar wind luminosity $L_w$  is obtained from the stellar content in the cluster; and \textit{iv)} the size of the Interstellar Bubble is deduced from optical observations and can be compared with the solutions outlined by \citet{Weaver}.

Free parameters and their limits are:\  \textit{i)} the conversion efficiency $\xi$ with $\xi < 1$; \textit{ii)} the electron-to-proton ratio $K_{ep}$ for which the range $[10^{-4},10^{-2}]$ can be reasonably used; the magnetic field in the \hii\ region that is at least equal to the mean Galactic field of $3\,\mu \mathrm{G}$ and  could hardly be higher than $100\,\mu \mathrm{G}$ \citep{Harvey-Smith2011}; \textit{iii)} the magnetic field turbulence level of $0.1 < \delta B/B < 1$; \textit{iv)} the index of turbulence of $\nu = 5/3$ or $3/2$;\textit{v)} the index of the injection spectrum is conservatively fixed to $s = 2$; and \textit{vi)} the coherence length of the turbulent magnetic field is fixed to $l_c=1$\,pc.

\section{Particle and photon spectra}
\label{chap3}
\begin{figure}[tp!]
\centering
\includegraphics[width=0.5\textwidth]{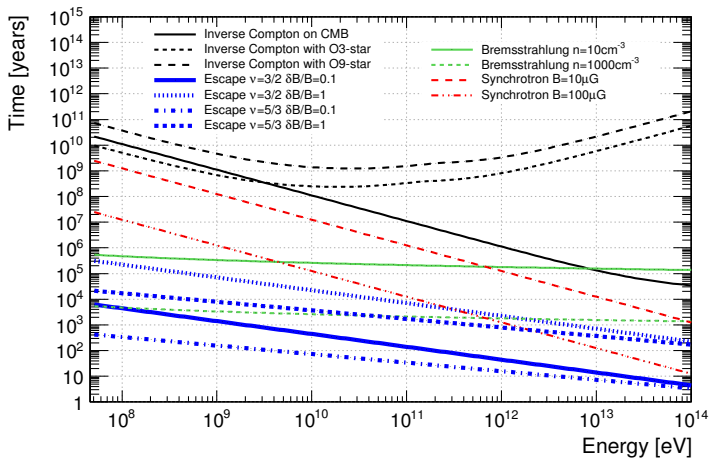}
\caption{Characteristic times for leptonic loss processes and escape assuming a typical cluster with an age of $1\,\mathrm{Myr}$ and a radius of the \hii\ region of $\RHii = 10\,\mathrm{pc}$.}
\label{fig:timeElec}
\end{figure}

\begin{figure}[tp!]
\includegraphics[width=0.5\textwidth]{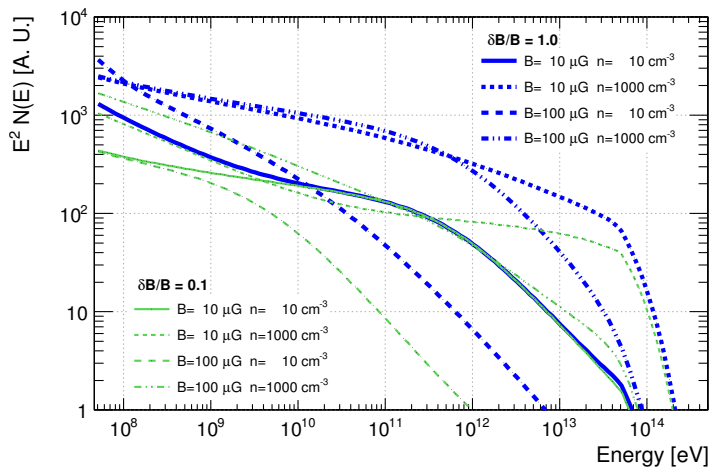}
\caption{Electron spectra assuming a typical cluster with an age of $1\,\mathrm{Myr}$ and with a radius of the \hii\ region of $\RHii = 10\,\mathrm{pc}$.
The electron injection spectrum has a spectral index of $s=2$ and an exponential energy cut-off set to $E_{e,max}=50$\,TeV. The index of turbulence is set to $\nu=3/2$ (Iroshnikov-Kraichnan turbulence).}
\label{fig:SpectrumElec}
\end{figure}

In this chapter the one-zone model presented in section~\ref{modelexplain} is applied to a generic star cluster  to understand the typical behaviour of star clusters. This generic cluster is $1\,\mathrm{Myr}$ old, contains one O3-type star, and is located in a \hii\ region with a radius of $\RHii=10\,\mathrm{pc}$ and densities of $\nHii = 10\,\mathrm{cm}^{-3}$, representing a low-density scenario, or $\nHii = 1000\,\mathrm{cm}^{-3}$, representing a high-density scenario. The electron and proton spectra and the subsequent \gr\ spectra are calculated.

\subsection{Electrons}

\begin{figure*}[tp!]
\centering
\includegraphics[width=\textwidth]{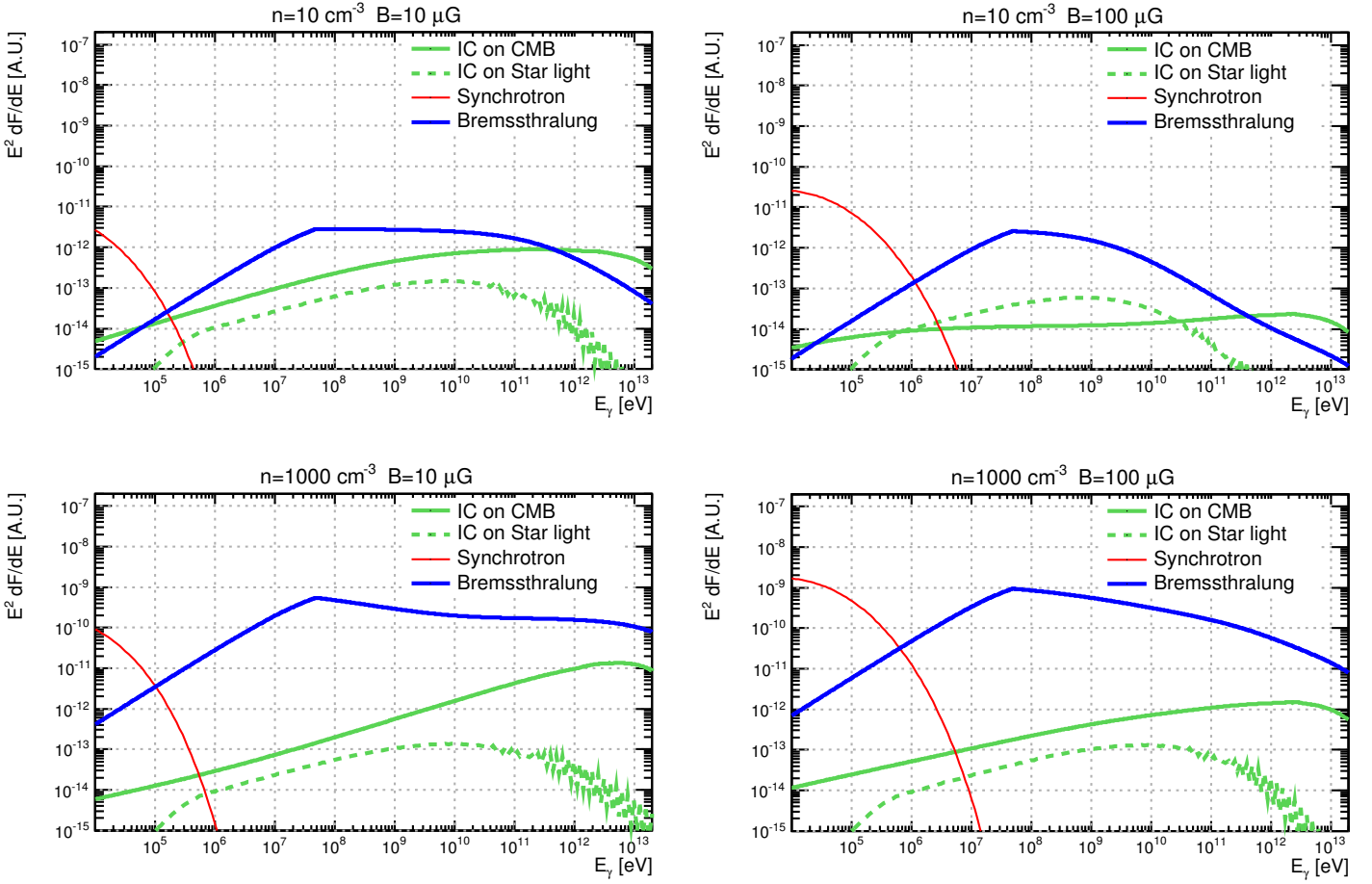}
\caption{Non-thermal emission for leptonic processes and different magnetic fields and densities assuming a typical cluster with an age of $1\,\mathrm{Myr}$ and radius of the \hii\ region of $\RHii = 10\,\mathrm{pc}$.
The spectral index of the injection spectrum is set to $s=2$, electron maximum energy is fixed to $E_{e,max}=50$\,TeV, turbulence index is $\nu=3/2$, magnetic turbulence is $\delta B/B=0.1,$ and electron-to-proton ratio is set to $K_{ep}=10^{-2}$.
}
\label{fig:GamLeptonic}
\end{figure*}

\subsubsection{Characteristic times}
As the electrons enter the \hii\ region, they suffer energy losses through several processes: IC scattering off the CMB, galactic photon fields, and star light from the cluster itself; synchrotron radiation in the magnetic field $B$ in the \hii\ region; and finally Bremsstrahlung depending on the density \nHii\ inside the \hii\ region.

The characteristic times of these processes are computed according to \citet{BLUMENTHAL} and are shown in Fig.~\ref{fig:timeElec}. They are compared to the escape times (eq.~\ref{tauesc}) for different values for the density, index of turbulence, and intensity and turbulence of the magnetic field. A physical process dominates if its characteristic time is the shortest.

In the low-density scenario, all processes are largely dominated by escape losses with the exception of the synchrotron emission in a high magnetic field, however, a high magnetic field is only expected at high densities \citep{Harvey-Smith2011}. In the high-density scenario the escape in the case of a magnetic field with a low level of turbulence is still dominant. Electro-magnetic losses only become significant or dominant in the case of high density and high level of magnetic turbulence: Bremsstrahlung losses become dominant at lower energies; synchrotron losses become significant or dominant in a high magnetic field at higher energies.

Low-energy electrons undergo IC scattering off the CMB and off star light from the cluster itself, but the Klein-Nishina effect becomes important for IC scattering off stellar radiation  at high energy (E>100\,GeV). The electrons scatter mainly on CMB. IC is still negligible when confronted with other processes.

These results call for two remarks. First, escapes dominate the other processes in most of the cases.  Second, at high energies the characteristic times of the dominant processes are well below the age of the cluster. These validate the assumption of a one-zone model with an uniform density of cosmic rays.

\subsubsection{Spectrum}

The different electron spectra resulting from a exponentially cut-off power-law injection with an index $s=2$ and a electron maximum energy $E_{e,max}=50$\,TeV provided to a generic cluster are shown in Fig.~\ref{fig:SpectrumElec}. Independent of the turbulence index several behaviours can be observed. Firstly, as the density increases, two antagonistic effects occur. On the one hand, the Bremsstrahlung losses increase, which changes the spectrum particularly at high energies; on the other hand, the injection is amplified, thereby increasing the spectrum normalization.
Secondly, at high energy, the spectrum drops as a result of synchrotron losses on the magnetic field. The drop occurs even at lower energies and is more marked as the magnetic field gets more intense.
Thirdly, the turbulence level has an impact when the synchrotron or Bremsstrahlung losses are significant, i.e. when the density and/or the magnetic field are high. The spectrum is then amplified at all energies.

\subsubsection{Non-thermal emission}

The photon fluxes are deducted from the electron spectra in the cluster and from the \gr\ emissivity of the different processes, according to \citet{BLUMENTHAL}. Figure~\ref{fig:GamLeptonic} presents the photon flux in the typical cluster, assuming $\delta B/B=0.1$ for two magnetic field intensities ($10\,\mu\mathrm{G}$ and $100\,\mu\mathrm{G}$) and two densities ($10\,\mathrm{cm}^{-3}$ and $1000\,\mathrm{cm}^{-3}$).

As expected, the synchrotron emission dominates the non-thermal emissions below 0.1\,GeV whatever the magnetic field and density. Then, Bremsstrahlung emission systematically dominates beyond 1\,GeV at high densities. At a low density, Bremsstrahlung emission dominates up to 1\,TeV and the IC emission takes over beyond. The IC emission generated from the star light remains generally low; the case here is the most optimistic since it corresponds to the photon flux generated with  the light of an O3-type star.

\begin{figure}[tp!]
\centering
\includegraphics[width=0.5\textwidth]{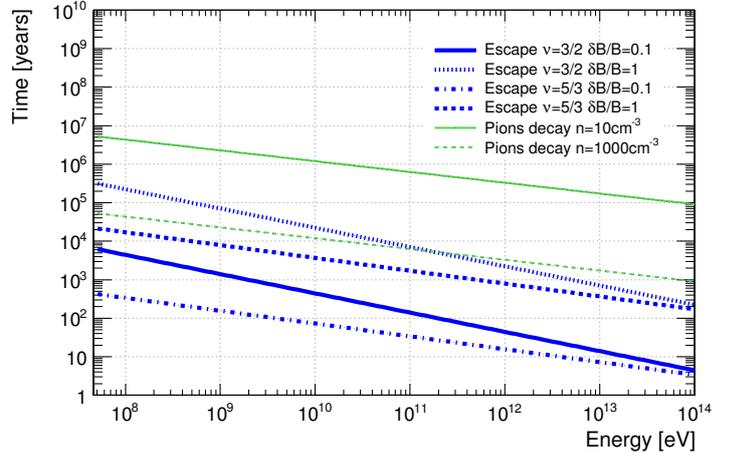}
\caption{Characteristic times for pion production and escape assuming a typical cluster with an age of $1\,\mathrm{Myr}$ and with a radius of the \hii\ region of $\RHii = 10\,\mathrm{pc}$.}
\label{fig:timeProton}
\end{figure}
\begin{figure}[tp!]
\includegraphics[width=0.5\textwidth]{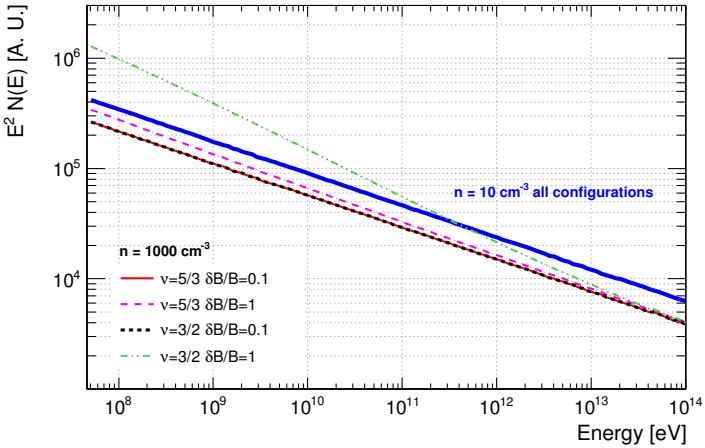}
\caption{Proton spectra assuming a typical cluster with an age of $1\,\mathrm{Myr}$ and radius of the \hii\ region of $\RHii = 10\,\mathrm{pc}$.
The proton injection spectrum has a spectral index of $s=2$. The blue line denotes the spectra for all configurations in the low-density scenario.}
\label{fig:SpectrumProton}
\end{figure}

\begin{figure*}[tp!]
\centering
\includegraphics[width=\textwidth]{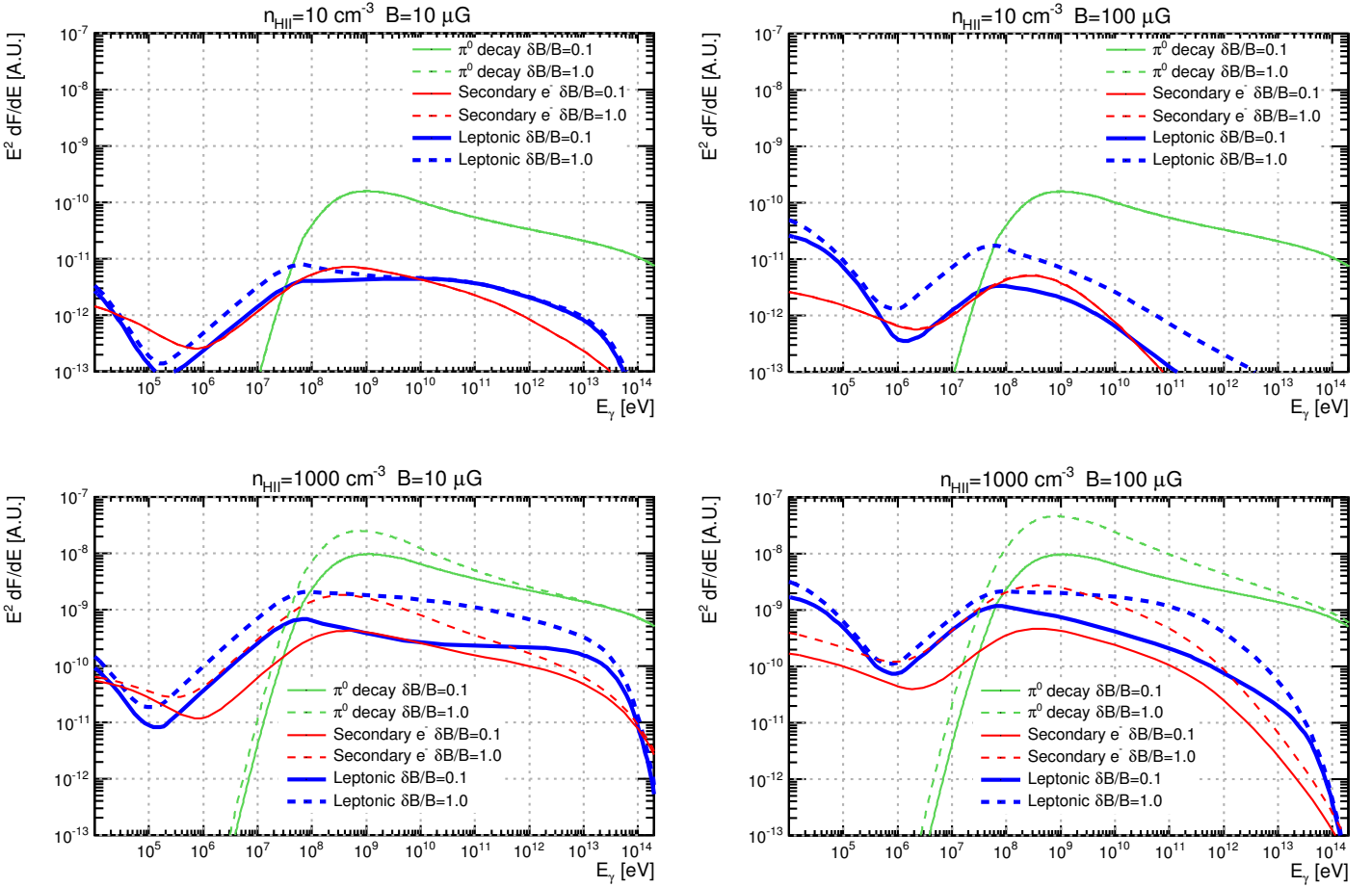}
\caption{Non-thermal emission from electrons and protons for different magnetic fields and densities assuming a typical cluster with an age of $1\,\mathrm{Myr}$ and radius of the \hii\ region of $\RHii = 10\,\mathrm{pc}$. The spectral index of the injection spectra is set to $s=2$, electron maximum energy is fixed to $E_{e,max}=50$\,TeV,  turbulence index is $\nu=3/2$, and electron-to-proton ratio is set to $K_{ep}=10^{-2}$.}
\label{fig:TotalEmission}
\end{figure*}

\subsection{Protons}
\subsubsection{Characteristic times}

For protons, the dominant loss process is pion production. Adiabatic losses do not play a role as the \hii\ region has an approximate stationary size, and the Coulomb interaction and ionization losses are negligible at the energies under consideration. Figure \ref{fig:timeProton} compares the characteristic times of pion decay to the escape time according to the density, index of turbulence, and magnetic field (intensity and turbulence level).

In most cases, pion production is dominated by escape losses beyond 1\,GeV. Pion production and escape losses are comparable in the case of high densities and high turbulence levels. Moreover, as one process clearly dominates the others, the proton spectrum should be monotonous.

\subsubsection{Spectrum}

The different proton spectra resulting from a power-law injection with an index $s=2$ provided to a generic cluster are shown in Fig.~\ref{fig:SpectrumProton}.

At low densities, the proton spectrum is only slightly sensitive to the index of turbulence and to the magnetic field turbulence level. It follows a power-law distribution with an index that is close to 2.3. At high densities, increasing turbulence level raises the proton spectrum at low energies. The spectrum softens and then reaches an index of 2.45. Slight differences appear between the two indices of turbulence.

Secondary electrons can be deducted from the proton spectra and from the emissivity of the charged pion production, according to \citet{2006PhRvD..74c4018K}. Their gamma-ray emissions are directly compared to those from primary electrons in the next part.

\subsection{Non-thermal emission from a stellar cluster \label{Nonthermalpredict}}

\subsubsection{Hadronic and leptonic emissions}

The simple one-zone model can predict non-thermal emission from both electrons and protons. Figure~\ref{fig:TotalEmission} presents the hadronic and leptonic emission in the typical cluster assuming an electron-to-proton ratio of $K_{ep} = 10^{-2}$.

These results call for three remarks. Firstly, at high energies, non-thermal emissions are clearly dominated by the pion decay because of the high density in these objects.  The leptonic emission can only dominate pion decay emission for a very optimistic $K_{ep}$ ratio greater than 10\%. Therefore, the assumption on the magnetic field does not influence the non-thermal emission above 1\,GeV. This result is robust considering parameters with their realistic limits (see \ref{paramresum}). Secondly, at low energies, the photon flux slightly depends on the intensity and turbulence level of the magnetic field and turbulence index. Thirdly, emissions from secondary electrons are comparable to an injection of primary electron with $K_{ep}=10^{-2}$. These emissions remain clearly dominated by $\pi^{0}$ decay representing several percent of hadronic emissions above 1\,GeV. In conclusion, the young star-forming regions are potentially hadronic sources of \grs.

\subsubsection{Temporal evolution}

\begin{figure}[tp!]
\centering
\includegraphics[width=0.5\textwidth]{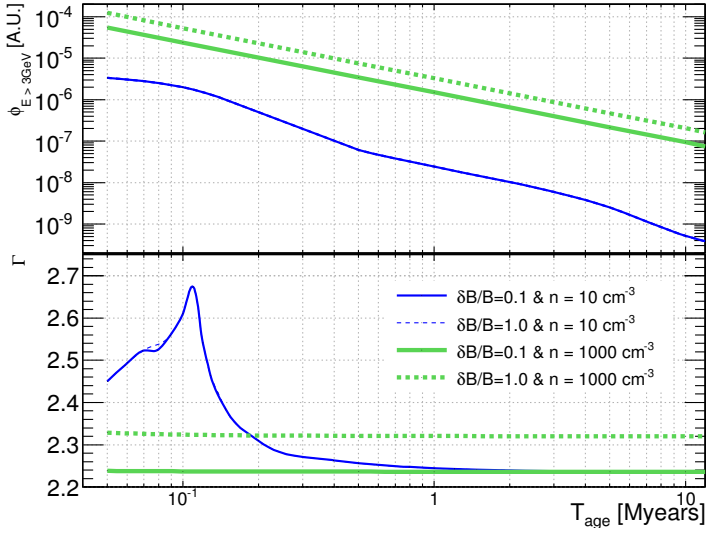}
\caption{Temporal evolution of the non-thermal emissions (spectral index and integrated flux above 3\,GeV) assuming a typical cluster with an age of $1\,\mathrm{Myr}$ and with a radius of the \hii\ region of $\RHii = 10\,\mathrm{pc}$.
The spectral index of the injection spectra is set to $s=2$, electron maximum energy is fixed to $E_{e,max}=50$\,TeV,  turbulence index is $\nu=3/2$, magnetic field strength is $B = 10\,\mu\mathrm{G,}$ and electron-to-proton ratio is set to $K_{ep}=10^{-2}$.}
\label{fig:Evolution}
\end{figure}

In this model, the luminosity is assumed to be constant over time. The bubble formed by the stellar winds grows with time. Equations \ref{inject} and \ref{inject2} show that the injection rate $Q_0$ is decreasing with time. Particles spectra and non-thermal emissions do not reach a steady state. The figure \ref{fig:Evolution} presents the temporal evolution of the integrated flux from $3$\,GeV and spectral index $\Gamma$ of the non-thermal emission, assuming a typical cluster. It can be seen that the integrated flux quickly reaches a maximum and then decreases with time following the injection model ($\phi_{E > 3GeV} \propto t^{-6/5}$). This flux is stronger if the density and/or the turbulence level are high. The spectral index quickly stabilizes if the density is high ($t_{stable}<50$\,kyr). In this case, this depends slightly  on the level of turbulence. At low densities, the spectral index typically  stabilizes after $10^{6}$ years and does not change with the turbulence level.

\section{Observational constraints}
\begin{table*}[tp!]
\caption{For each cluster, this table summarizes the important astrophysical parameters of the model.}
\centering
\begin{tabular}{ l | c c c c c c c | r r | c}
\hline
\hline

Cluster    & O stars        &Luminosity               & \RHii\  & $R_{b}$ & \nHii\
            & Log(Age)    &  D    &  \multicolumn{2}{c|}{\fermi\ analysis}  & $\xi_\mathrm{max}$   \\
name       & (most massive) &($\mathrm{erg\,s}^{-1}$) & (pc)    & (pc)    & ($\mathrm{cm}^{-3}$)
            & Log(years)  & (kpc) &  TS & $\Phi^{95\%}_{UL}$ ($\mathrm{cm}^{-2}\mathrm{s}^{-1}$) & (\%) \\
\hline
\hline
NGC\,2244    &    4 (O4)
            & $1.0\times 10^{37}$
            &   16.9
            & 6.2
            &  15
            & 6.28
            &  1.55
            &  22.1
            &   $4.61\times10^{-10}$
            & 5.80\\
NGC\,1976    &    4 (O7)
            & $1.5\times 10^{36}$
            &   3.7
            &  2
            &  8900
            & 6.4
            & 0.4
            &
            5.6
            &   $3.50\times10^{-10}$
                & 6.69\\
\hline
NGC\,2175    &    1 (O6.5)
            & $1.3\times 10^{36}$
            &  12
            & 2.56$^{*}$
            &  13
            & 6.3
            &  2.2
            &   7.1
            &   $2.14\times10^{-10}$
            &  9.81\\
NGC\,3324    &    2 (O6.5)
            & $1.6\times 10^{36}$
            &  6.5
            & 2.32$^{*}$
            &  33
            & 6.4
            &  3
            &  2.7
            &   $5.54\times10^{-10}$
            &  100\\
RCW\,8       &    2 (O8.5)
            & $4.8\times 10^{35}$&  2.2
            & 1.86$^{*}$  &  91
            & 6.78  &  4.2
            &  11.2
            &   $1.75\times10^{-10}$
            &  100\\
RCW\,62      &    10 (O6)
            & $9.2\times 10^{36}$
            &  25.6     & 2.59$^{*}$
            &  430   & 6.8   &  2.2
            &   0.1
            &   $3.54\times10^{-10}$
            &  0.13\\
NGC\,6618    &    17 (O4)
            & $3.3\times 10^{37}$
            &   4       & 1.69$^{*}$
            &  470   & 6     &  1.6
            &   1.4                  &  $4.20\times10^{-11}$
            &  0.27 \\
NGC\,2467    &    3 (O3)
            & $1.4\times 10^{37}$
            &   4
            & 1.51$^{*}$
            &  550
            & 6.3
            &  4.1
            &   3.2
            &   $9.57\times10^{-11}$
            &  7.06\\
\hline
\hline
\end{tabular}
 \tablefoot{The radius \RHii\  of the \hii\ region and the density \nHii\ , which are fixed by optical observations, the age of the cluster and its distance. The luminosity is computed from the composition of O-stars, according to \citet{1990A&A23134N} and \citet{Massey}. The bubble radii indicated with a ${*}$ were computed using equation 12 from \citet{Weaver}.
 The second part of the table presents the results of the \fermi\ analysis. TS is the test statistic value for \gr\ emission at the position of the cluster and $\Phi^{95\%}_{UL}$ is the upper limit on the integrated \gr\ flux at a 95\% confidence level. Finally, $\xi_\mathrm{max}$ gives the upper limit of the stellar wind luminosity converted into particle acceleration, obtained with the one-zone model.}
\label{BestClust}
\end{table*}

\subsection{Sample of stellar clusters \label{ClusterSelection}}

We selected a sample of clusters
embedded in \hii\ regions from the Galactic O-Star Catalog
\footnote{\url{http://ssg.iaa.es/en/content/galactic-o-star-catalog}}
\citep[GOSC;][]{GOSC}
  version 2.4, containing 1285 O-type stars,  following two main criteria:

\begin{itemize}
  \item The cluster must correspond to the described model:\ \textit{i)} the cluster is young with an age of less than 10\,Myr; \textit{ii)} no evolved star is present in the cluster; \textit{iii)} No supernova explosion has occurred in the cluster yet; and \textit{iv)} the shape of the \hii\ region is almost spherical.
  \item The main astrophysical properties of the cluster must be known:\ \textit{i)} the population of O-type stars; \textit{ii)} the radius and density of the \hii\ region; and \textit{iii)} the distance to the cluster.
\end{itemize}

The eight clusters corresponding to these criteria are summarized in Table~\ref{BestClust}.

\subsubsection{Astrophysical properties}

The size of the bubble is known in only two clusters: NGC\,2244 in the Rosette Nebula and NGC\,1976 in the Orion Nebula. The results obtained from these two objects are discussed in detail. These two well-known objects have very different mean densities: $15\,\mathrm{cm}^{-3}$ in the case of the Rosette Nebula and $8900\,\mathrm{cm}^{-3}$ in the case of the Orion Nebula. They are, therefore, ideal to test the model described in this paper.

The star cluster NGC\,2244, belonging to the Mon OB2 association, has an age of about $1.9\times10^6$ years \citep{2002AJ123892P} and contains six O-type stars.
The cluster releases $10^{37}\,\mathrm{erg\,s}^{-1}$ in stellar winds, creating a giant Interstellar Bubble ($R_{b}=6.2\,\mathrm{pc}$) of rarefied matter in the Rosette Nebula. The Rosette Nebula is a \hii\ region of more than $10^{5}$ solar masses \citep{0004-637X-476-1-166}, which is ionized by the emission of its embedded star cluster. The nebula is located at a distance of 1.55\,kpc \citep{refId0} in the Monoceros constellation and has  an apparent size of 1.3\dgr.

The Trapezium cluster (NGC\,1976) is located at a distance of 0.41\,kpc \citep{2009ApJ...700..137R} and belongs to the Ori OB1d association. It contains three O-type stars.
The mechanical energy released by the cluster of more than $10^{36}\,\mathrm{erg\,s}^{-1}$ creates a bubble of rarefied matter with a radius of $R_{b}=2\,\mathrm{pc}$  and ionizes the very dense ($n = 8900\,\mathrm{cm}^{-3}$) \hii\ region of the Orion Nebula.

 The stellar content is not found in the GOSC for two clusters.
RCW\,8 contains at least two O-type stars, O8.5V and O9.5V \citep{RCW8StellarContent}. The Omega Nebula (NGC\,6618) is one of the most luminous \hii\ regions in our Galaxy. It contains 17 O-type stars \citep{OmegaStellarContent}, with a binary O4+04 system CEN\,1. A non-thermal radio emission was observed \citep{2009RMxAA..45..273R} and a power-law X-ray emission \citep{2007ApJS..169..353B} in NGC 6618.

The properties of the other clusters are summarized in the left part of Table~\ref{BestClust}. Their luminosity is computed from the O-star composition according to \citet{1990A&A23134N} and \citet{Massey}. The most massive stars clearly dominate the total luminosity of the star clusters.  We calculated the total luminosity using equation 12 of \citet{Weaver} for  clusters where the size of the bubble is unknown.

\subsection{Fermi data analysis}

\begin{figure*}[p]
\centering
\includegraphics[width=6.3cm]{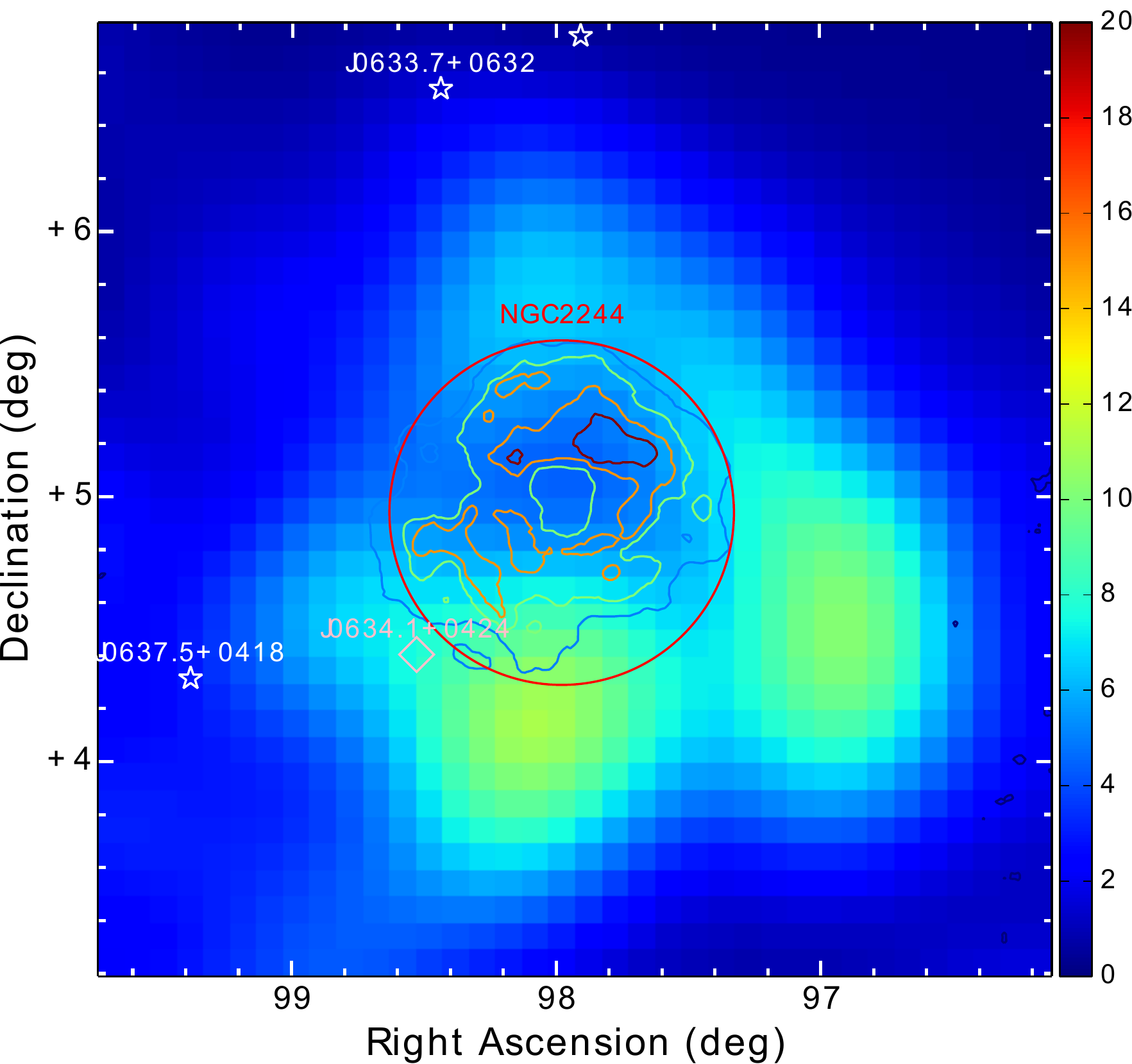}
\includegraphics[width=6.3cm]{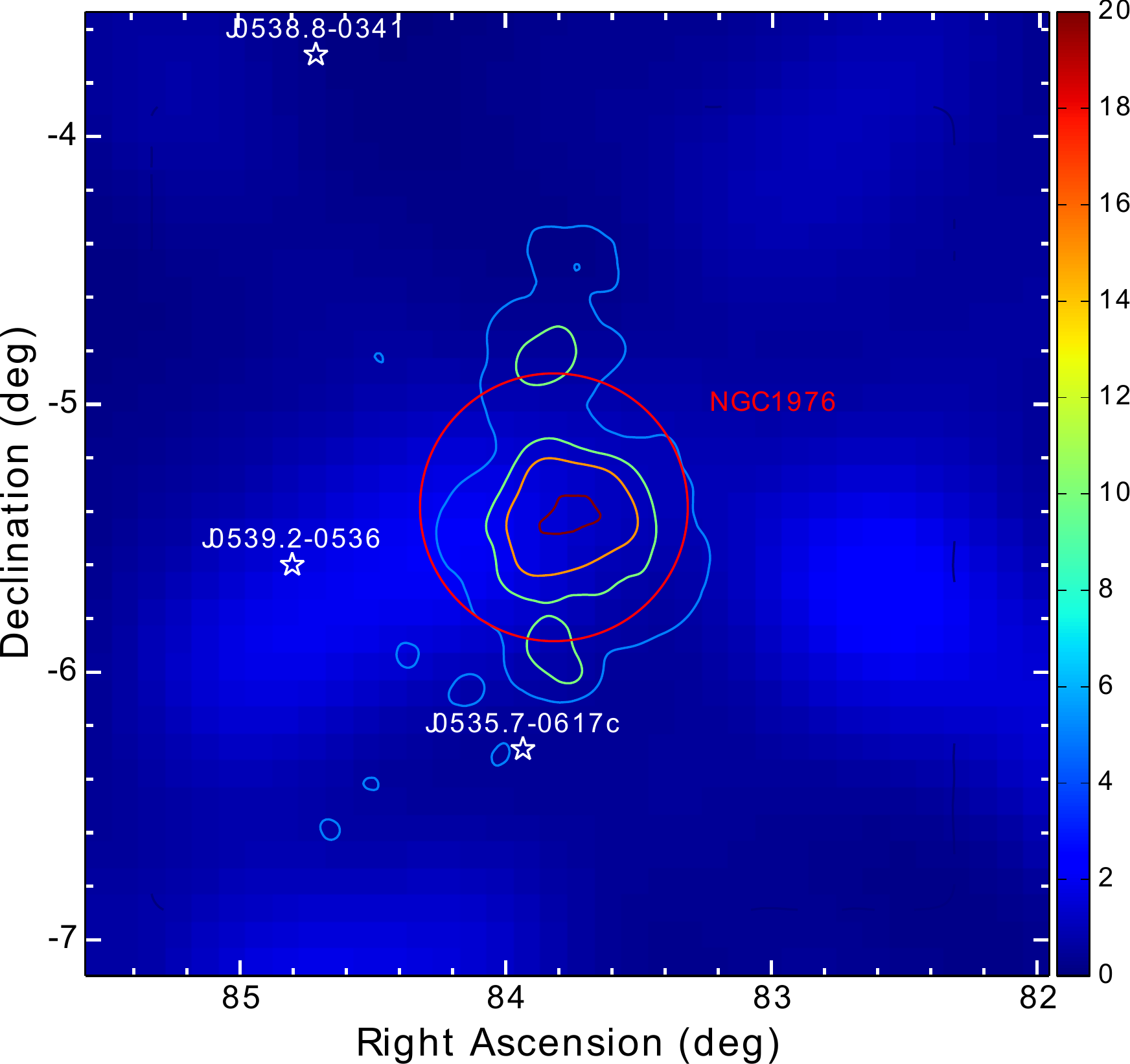}
\includegraphics[width=6.3cm]{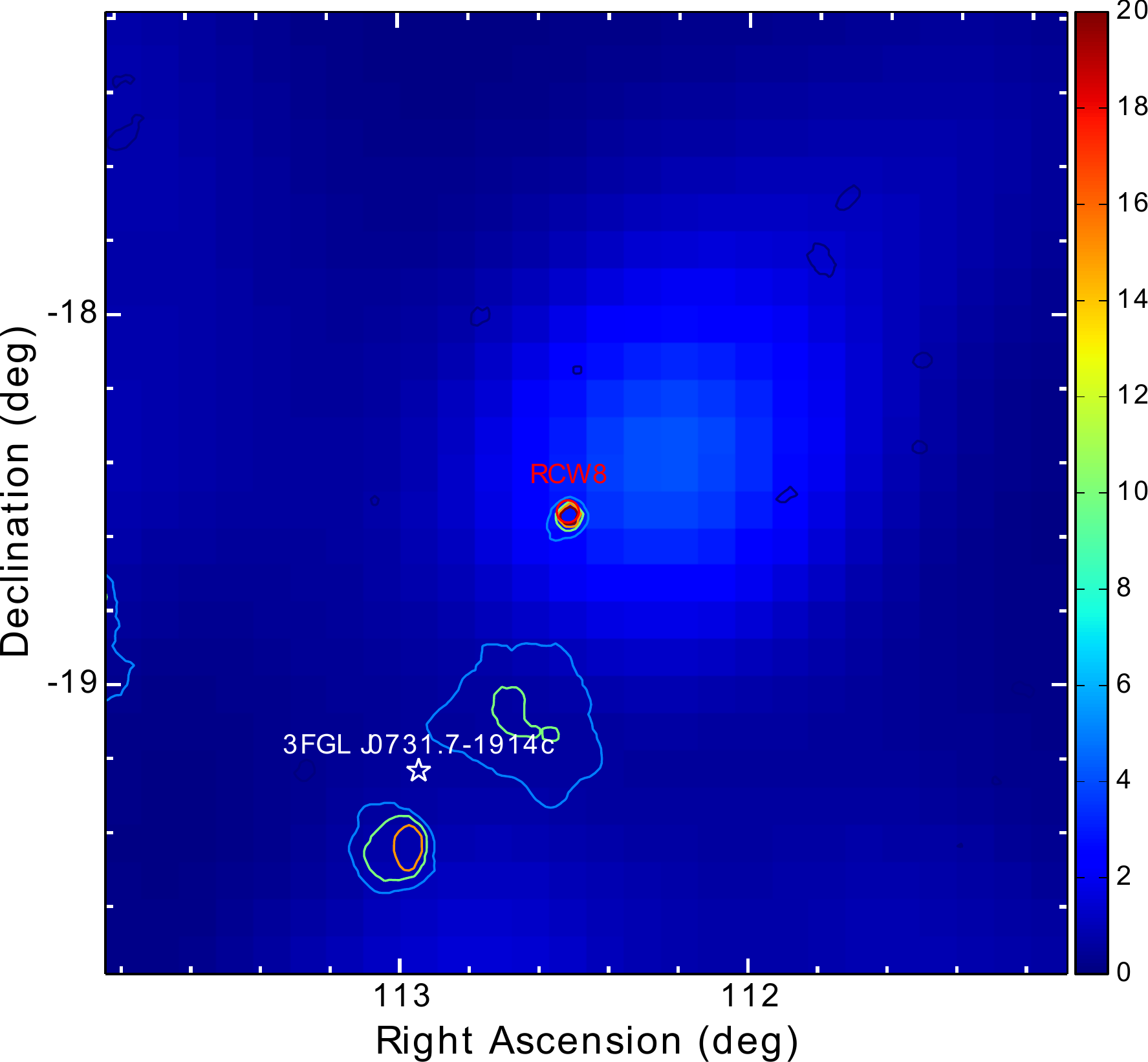}
\includegraphics[width=6.3cm]{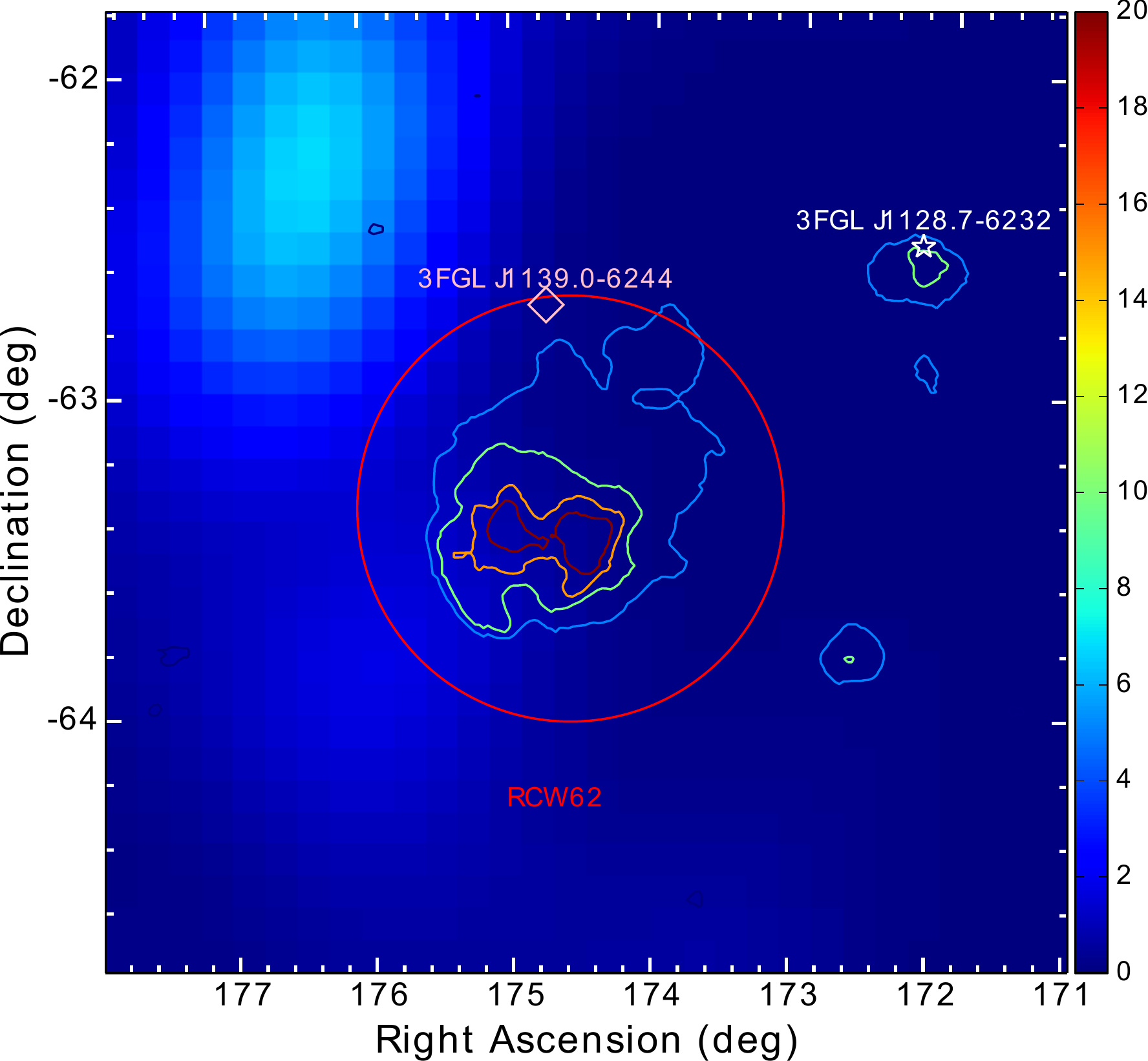}
\includegraphics[width=6.3cm]{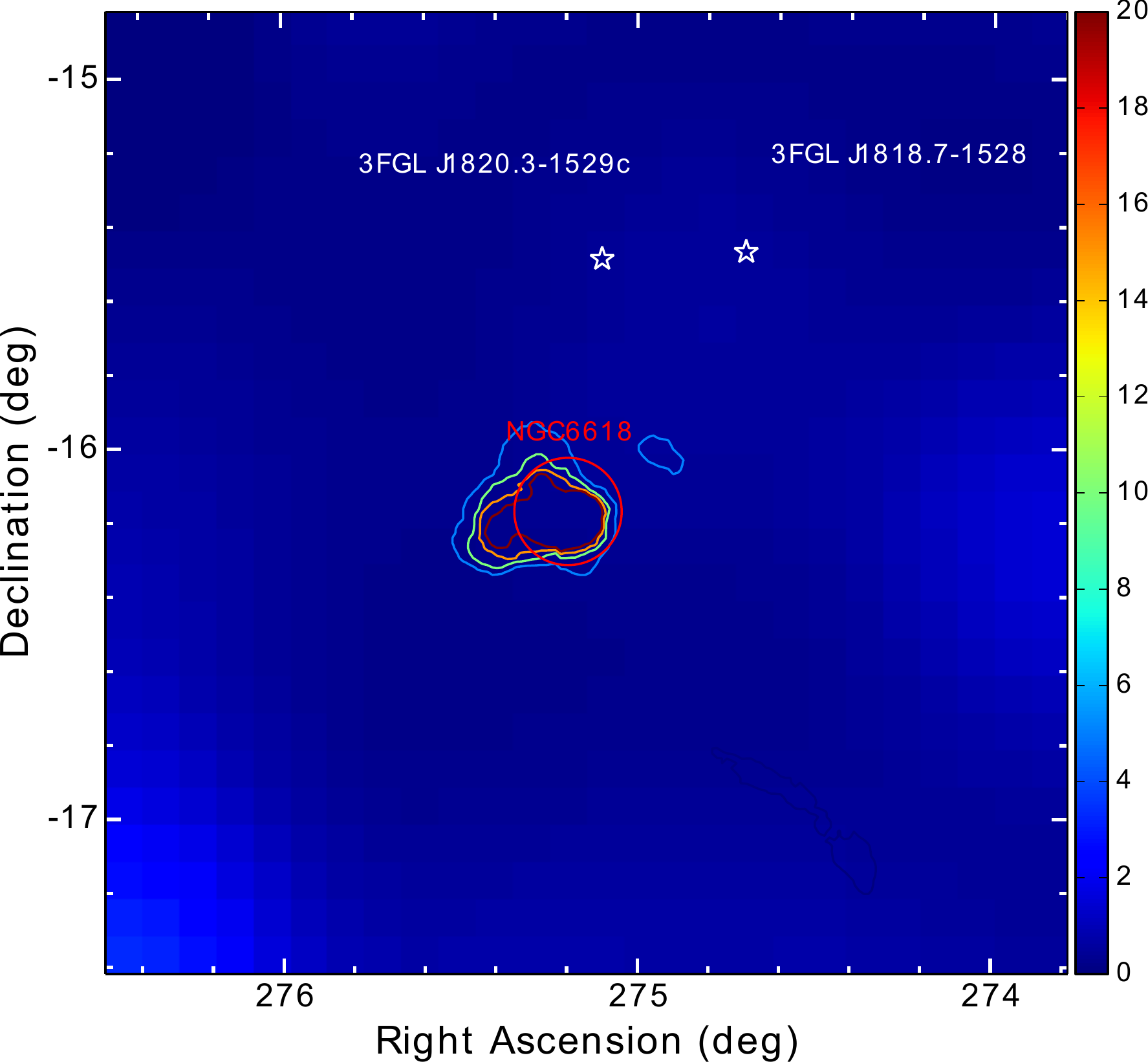}
\includegraphics[width=6.3cm]{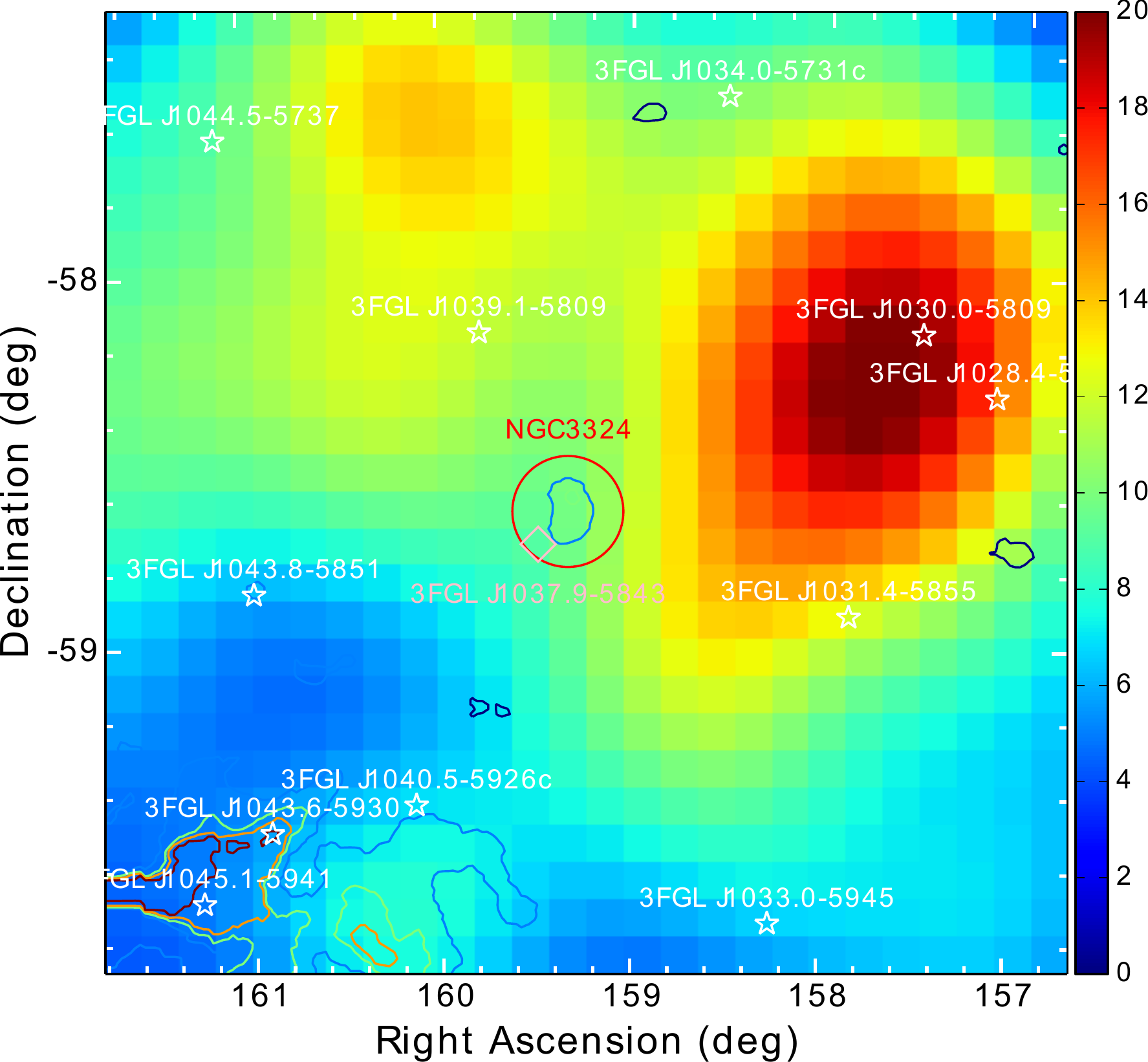}
\includegraphics[width=6.3cm]{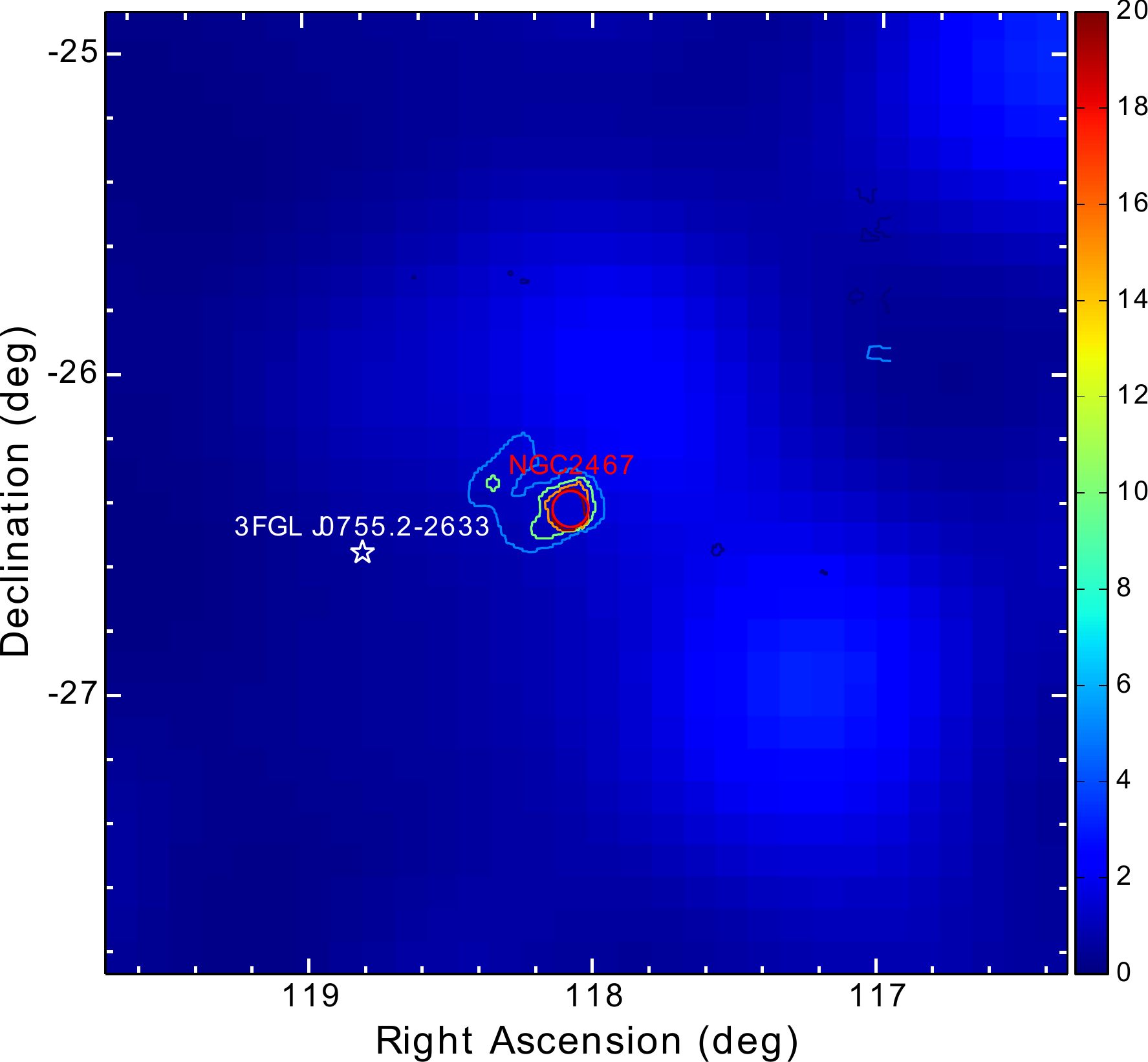}
\includegraphics[width=6.3cm]{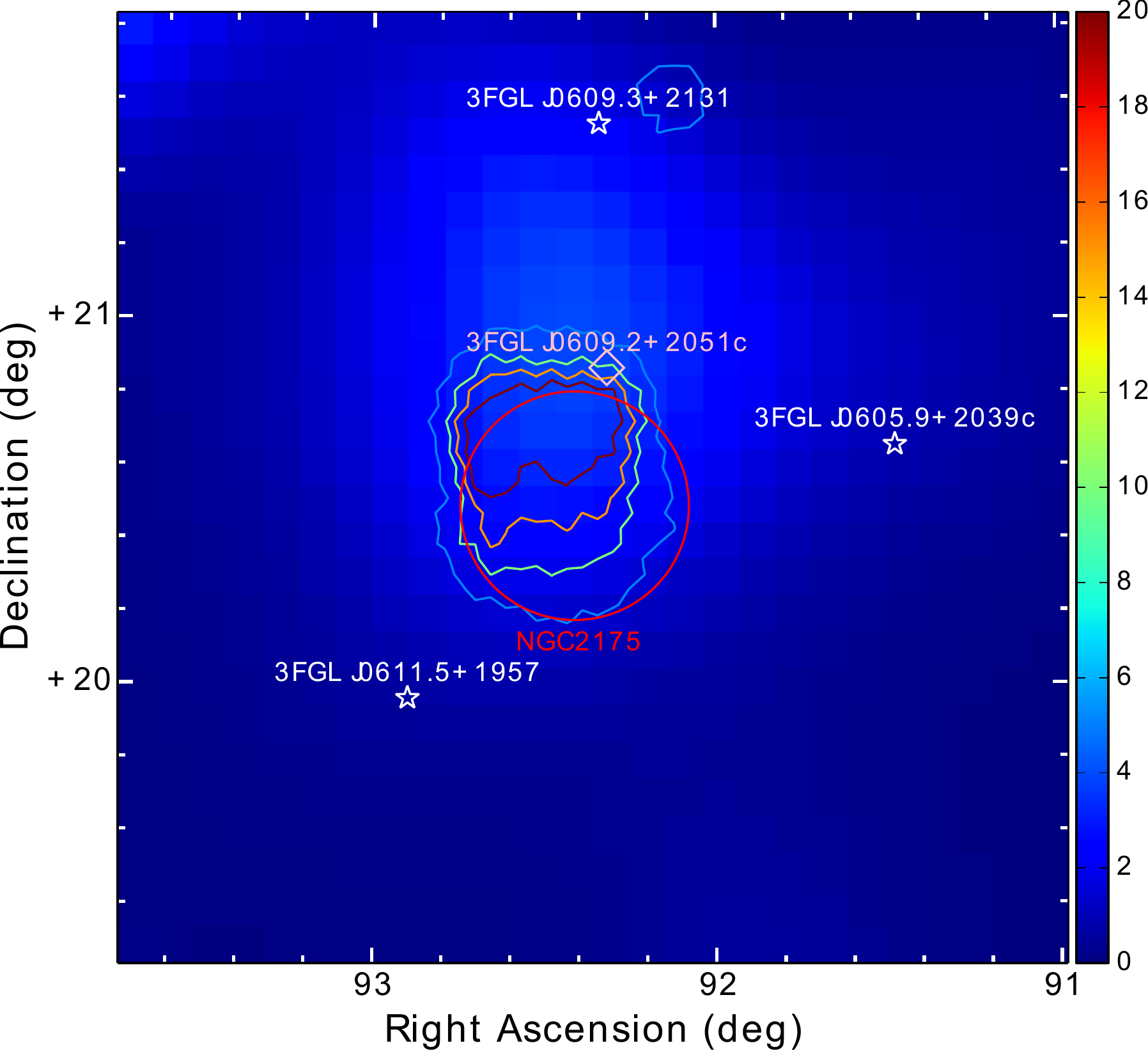}
\caption{Test statistics (TS) maps of the regions around the star clusters. The red circle denotes the region of interest used in the analysis; white stars (and pink diamonds) represent sources from the 3FGL catalogue with free (fixed) spectral parameters. Contours represent the \hii\ regions \citep{2001PASP..113.1326G}.}
\label{fig:TSmaps}
\end{figure*}

Gamma-ray emission from the selected clusters is searched for in the data from the Large Area Telescope (LAT) on board the \emph{Fermi} satellite. The \fermi\ is a pair-conversion telescope that is sensitive to photons with energies from 20\,MeV to more than 300\,GeV \citep{2009ApJ...697.1071A}. The data used in this work were recorded  between 2008 August 4 and 2015 June 15 and are reconstructed with the \emph{Pass 8} reprocessing chain \citep{2013arXiv1303.3514A}. The analysis is performed using the Fermi {\tt ScienceTools} v10r0p5 software with the python interface {\tt enrico} \citep{2013arXiv13074534S}.

Only events belonging to the \emph{source} event class and \emph{front+back} event type are used. All events within a circular region of interest (ROI) of 10\dgr radius centred on the source position are extracted. A zenith angle greater than 90\dgr and time intervals when the rocking angle is larger than 52\dgr were excluded from the analysis to reduce the $\gamma$-bright limb of the Earth. Finally, only events with an energy between 3\,GeV and 300\,GeV are selected.


The unbinned maximum-likelihood method implemented in the {\tt gtlike} tool is performed to compute the spectral analysis with the {\tt P8R2\_SOURCE\_V6} instrument response functions (IRFs).
The star cluster emission is modelled as a diffuse source corresponding to a projected, uniformly filled sphere with a radius \RHii\  and a power-law energy spectrum (powerlaw2).

The background is modelled using the  \gr\ sources of the third \fermi\ source catalogue \citep[3FGL;][]{3FGLFermi-LAT} based on four years of observations.
Sources of the 3FGL catalogue that coincide with the star clusters are deleted from the background model to account for the case that these objects are part of the emission from the star cluster.
The spectral parameters of the sources outside the star cluster but within 3\dgr\ of the ROI are kept free, and the spectral parameters of the sources further away are fixed to the values from the 3FGL catalogue.
The background model also contains contributions from the Galactic diffuse emission (gll\_iem\_v06) and from the isotropic background (iso\_P8R2\_SOURCE\_V6\_v06).
In a first step the normalization of the diffuse emissions are kept free in the fit  to calculate the best-fit test statistics and to generate sky maps.

In the absence of a signal (test statistics smaller than 25), an upper limit on the integrated \gr\ flux between 3\,GeV and 300\,GeV is calculated using the Integral method implemented in the \fermi\ ScienceTools and assuming a power law with a spectral index of  $\Gamma$=2.25 predicted by the one-zone model (\ref{Nonthermalpredict}). In order to obtain a conservative upper limit, the normalizations of the Galactic diffuse emission model is reduced by 6\% of the best-fit values for the calculation of the upper limits.

\subsection*{NGC\,2244 - Rosette Nebula}

The background model for the Rosette Nebula contains four point-like sources within 3\dgr of NGC\,2244 for which the spectral parameters are kept free in the fit. Another18 point-like sources, which are located further away the spectral parameters, are fixed to the values of the 3FGL catalogue. Possible contamination at lower energies from \gr\ emission, produced by the shock from the supernova remnant Monoceros Loop, interacting with the \hii\ region of the Rosette Nebula \citep{2014cosp40E1419K} is avoided with the elevated lower energy cut of 3\,GeV. The point-like source 3FGL\,J0634.1+0424 is located within 0.6\dgr of NGC\,2244 and is deleted from the background model. This source has a steep and curved spectrum described by a LogParabola model. Therefore, it will not significantly contribute to the \gr\ emission above 3\,GeV.

No significant \gr-emission from NGC\,2244 and the Rosette Nebula is detected in this analysis; the test statistics for this source is TS=17.5. The first panel of Fig.~\ref{fig:TSmaps} shows the test statistics map of the surroundings of the Rosette Nebula. The upper limit on the \gr\ flux  above 3\,GeV at 95\% confidence level is $\Phi_\gamma<5.07\times10^{-10}$cm$^{-2}$s$^{-1}$.

\subsection*{NGC\,1976 - Orion Nebula}

The background model of the Orion Nebula contains three point-like sources within 3\dgr with free parameters and another 15 point-like sources further away with fixed spectral parameters. The second panel of Fig.~\ref{fig:TSmaps} shows the test statistics map of the surroundings of the Orion Nebula. The Orion Nebula itself is not detected with a test statistics of TS=5.6. The upper limit (95\% confidence level) on the \gr\ flux above 3\,GeV is $\Phi_\gamma<3.80\times10^{-10}$cm$^{-2}$s$^{-1}$.

\subsection*{Others clusters}

The analyses of the remaining clusters are conducted in the same way: 3FGL sources coinciding with the star cluster (which are 3FGL\,J1037.9-5843 in the case of NGC\,3324, 3FGL\,J1139.0-6244 in the case of RCW\,62, and 3FGL\,J0609.2+2051c in the case of NGC\,2175) are deleted from the background model, the spectral parameters of sources within 3\dgr are kept free in the fit and the spectral parameters of sources further away are fixed to the 3FGL values.

No significant emission from any of these clusters is detected. The test statistics maps of all clusters are shown in Fig.~\ref{fig:TSmaps}. The test statistics of the clusters and the corresponding upper limits on the \gr\ flux are summarized in Table~\ref{BestClust}.

\section{Data-model comparison}
\begin{figure}[tp!]
  \centering
  \includegraphics[width=0.5\textwidth]{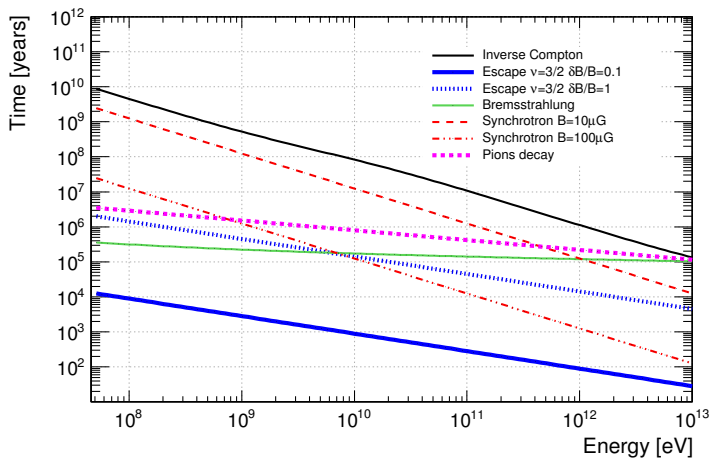}
     \caption{Characteristic times for escape; leptonic and hadronic loss processes in the \textbf{Rosette Nebula} (age=1.9$\,$My, $\RHii=16.9\,\mathrm{pc}$, $\nHii=15\,\mathrm{cm}^{-3}$).}
        \label{Fig:Rosette_Time}
  \end{figure}

\begin{figure}[tp!]
\centering
\includegraphics[width=0.5\textwidth]{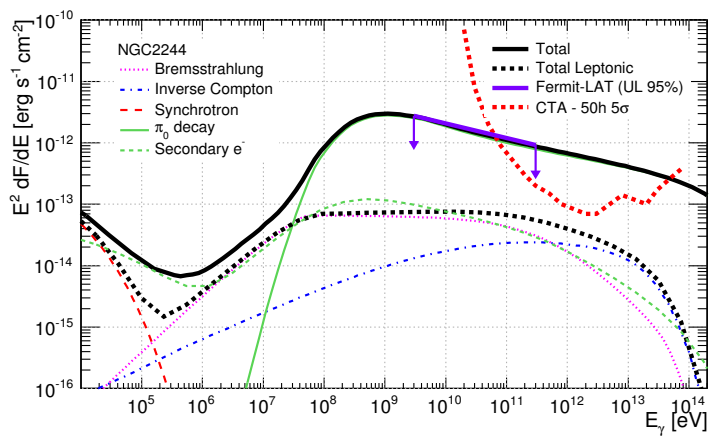}
\caption{\fermi\ upper limits and theoretical expectations on the spectral energy density deduced from models of the \textbf{Rosette Nebula} for realistic parameters: the turbulence index is $\nu=3/2$, magnetic turbulence is $\delta B/B=0.1,$ and electron-to-proton ratio is set to $K_{ep}=10^{-2}$. The magnetic field is $10\,\mu\mathrm{G}$. The red dashed line represents the sensitivity of CTA ($5\,\sigma$ detection in 50\,h, \cite{2012AIPC1505769B}).}
\label{fig:ModelFitsRosette}
\end{figure}

The non-detection of \gr\ emission with \fermi\ and the upper limits on the integrated \gr\ flux can be used to put constraints on the particle acceleration in the cluster. The goal is to determine the fraction $\xi$  of the stellar wind luminosity converted into accelerated particles in each cluster.

In the one-zone model discussed here \gr\ emission is produced by cluster-accelerated, relativistic particles interacting with the dense material in the surrounding \hii\ region. As shown in section \ref{chap3}, the \gr\ emission is clearly dominated by pion decay from the \hii\ region. Thus, the density \nHii\ of the \hii\ region is the crucial parameter. The second important parameter is the magnetic field that impacts both protons and electrons through diffusion and electrons through synchrotron radiation. As shown by \citet{Harvey-Smith2011}, however, the magnetic field in the \hii\ region directly depends on the matter density \nHii. Therefore, the two well-studied clusters, NGC\,2244 and NGC\,1976, which have very different densities \nHii, are ideal candidates to study particle acceleration and \gr\ emission of stellar clusters in detail.

\subsection{NGC\,2244 - Rosette Nebula}

With a density of $\nHii = 15\,\mathrm{cm}^{-3}$, in the \hii\ region NGC\,2244 and the Rosette Nebula can be considered as a prime example for a low-density scenario.

Figure~\ref{Fig:Rosette_Time} shows the characteristic times for the different energy loss processes for the Rosette Nebula. Above 1\,GeV the escape timescale in the case of low magnetic field perturbations (solid blue line) is shorter than any other process. Only synchrotron emission in a strong magnetic field (dot-dashed red line), which is unlikely owing to the low density in the Rosette Nebula, dominates the escape time at higher energies in the case of high magnetic perturbations (dashed blue line). As pion decay is the most dominant process after escape, the impact of synchrotron losses on the \gr\ emission can be neglected.

The spectral energy distribution of the predicted emission for a specific set of parameters ($K_{ep} = 10^{-2}$, $\delta B/B = 0.1$, $B = 10\mu G$, $\nu = 3/2$) is shown in Fig.~\ref{fig:ModelFitsRosette}. The upper limit on the \gr\ flux constrains the ratio of the stellar wind luminosity that is converted into accelerated particles, in this particular case, to $\xi < 5.5$\%. Exploring the possible set of the parameters $K_{ep}$, $\delta B/B$ and $\nu$ according to \ref{paramresum} demonstrates that not more than a fraction $\xi_\mathrm{max} = 5.8$\% of the stellar wind luminosity of NGC\,2244 can be converted into relativistic particles.

\begin{figure}[tp!]
  \centering
  \includegraphics[width=0.5\textwidth]{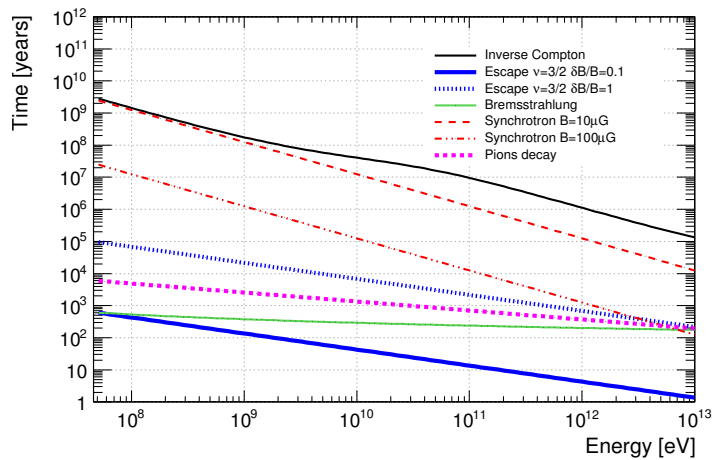}
     \caption{Characteristic times for escape, leptonic, and hadronic loss processes in the \textbf{Orion Nebula} (age=2.5$\,$My, $\RHii=3.7\,\mathrm{pc}$, $\nHii = 8900\,\mathrm{cm}^{-3}$).}
        \label{Fig:orion_Time}
  \end{figure}

\subsection{NGC\,1976 - Orion Nebula}

With a density of $\nHii = 8900\,\mathrm{cm}^{-3}$, NGC\,1976 can be considered as a prime example for a high-density scenario.

Figure~\ref{Fig:orion_Time} shows the characteristic times for the Orion Nebula. Contrary to the Rosette Nebula, Bremsstrahlung (green line) and pion production (dashed magenta line) become dominant. They even dominate the escape timescale at high turbulence levels (dashed blue line).

\begin{figure}[tp!]
\centering
\includegraphics[width=0.5\textwidth]{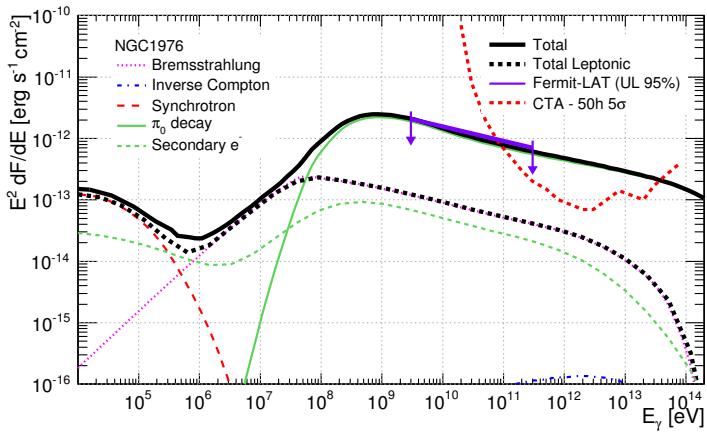}
\caption{\fermi\ upper limits and theoretical expectations on the spectral energy density deduced from models of the \textbf{Orion Nebula} for realistic parameters: the turbulence index is $\nu=3/2$, magnetic turbulence is $\delta B/B=0.1,$ and electron-to-proton ratio is set to $K_{ep}=10^{-2}$. The magnetic field is $100\,\mu\mathrm{G}$. The red dashed line represents the sensitivity of CTA ($5\,\sigma$ detection in 50\,h; \cite{2012AIPC1505769B}).}
\label{fig:ModelFitsOrion}
\end{figure}

The spectral energy distribution of the \gr\ emission for the parameters $K_{ep} = 10^{-2}$, $\delta B/B = 0.1$, $B = 100\mu G$ and $\nu = 3/2$ is shown in Fig.~\ref{fig:ModelFitsOrion}. The upper limit on the \gr\ emission restricts the conversion ratio of stellar wind luminosity to accelerated particles, in this case, to  $\xi < 6.3$\%. Exploring the possible parameter space (according to \ref{paramresum} part) demonstrates that not more than  $\xi_\mathrm{max} = 6.7$ \% of the stellar wind luminosity of NGC\,1976 can be converted into relativistic particles.

\subsection{Sample of embedded star clusters}

The maximum fractions of conversion of stellar luminosity into relativistic particles for all the clusters are summarized in Table~\ref{BestClust}.
The conversion fraction $\xi_\mathrm{max}$ for clusters  NGC\,2175 and NGC\,2467 are around 10\%, which is similar to the Rosette and Orion Nebulae discussed above.


The clusters RCW\,62 and NGC\,6618 set very low limits on $\xi_\mathrm{max}$ of less than 1\%. With similar luminosities and distances as NGC\,2244, the much higher densities in the \hii\ regions allow us to probe much lower levels of relativistic protons in the cluster.

The upper limits on the \gr\ flux obtained on NGC\,3324 and RCW\,8 do not allow us to reach any conclusion on the luminosity conversion into relativistic particles. This is because of the low stellar wind luminosity (created by only two O-type stars of relatively low mass), large distance (as they are the most and third-most distant objects in the sample), and relatively low density \nHii\ in the \hii\ region.

\section{Discussion}
\subsection{Influence of the main Parameters}
\label{influence}
So far, we tested the maximum conversion fraction of stellar wind luminosity into relativistic particles for a magnetic turbulence level of $\delta B/B = 0.1$, coherence length of $l_c = 1\,\mathrm{pc,}$ and spectral index of the injection spectrum of $s = 2$ (see section \ref{paramresum} for details on these parameters). Now the influence of these parameters on the maximum conversion fraction $\xi_\mathrm{max}$ is tested for the star clusters NGC\,2244 and NGC\,1976  to verify the reliability of the presented results. The impact of these parameters on the maximum conversion fraction is summarized in Table~\ref{Influence}.

\begin{table}[tp!]
  \caption{Influence of the main parameters (magnetic turbulence level $\delta B/B$, coherence length $l_c$, index of injection $s$ with $E_\mathrm{max}=10\,$PeV) on the upper limit of the fraction of mechanical energy converted into accelerated particles $\xi_\mathrm{max}$ for NGC\,2244 (Rosette Nebula) and NGC\,1976 (Orion Nebula).}
\centering
\begin{tabular}{ l |c c | c c | c c}
\hline
\hline
     &\multicolumn{6}{c}{$\xi_\mathrm{max} (\%)$}    \\

\hline

    &\multicolumn{2}{|c|}{$\delta B/B$}    & \multicolumn{2}{c|}{$l_c$} & \multicolumn{2}{c}{$s$} \\
Cluster       & $10^{-2}$&$10^{2}$&0.5\,pc& 2.0\,pc&1.5&2.5\\
\hline
\hline
NGC\,2244      & 5.95 & 5.10 & 5.88 & 5.88 & 16.06 & 0.85  \\
NGC\,1976      & 7.53 & 1.28 & 6.29 & 7.34 & 23.12 & 0.90 \\
\hline
\hline
\end{tabular}
\label{Influence}
\end{table}

With a decreasing magnetic turbulence level $\delta B/B$ or increasing coherence length $l_c$ diffusion losses become more important. Therefore, the expected \gr\ flux is lower. In turn, the maximum conversion efficiency $\xi_\mathrm{max}$ is higher. This trend is even more important for the high-density scenario, such as in NGC\,1976. However, the obtained $\xi_\mathrm{max}$ are lower than 10\%. The results presented here are robust in the change of the parameters $\delta B/B$ and $l_c$.

The influence of the spectral index $s$ of the injection spectrum is more important. A change of the injection index changes the spectral index $\Gamma$ of the \gr\ emission. For spectral indices of $s=1.5$ and $s=2.5$, the spectral index of the \gr\ emission is $\Gamma = 1.7$ and $\Gamma = 2.7$, respectively. We repeated the analysis of the \fermi\ data to obtain upper limits on the \gr\ flux for these indices. With increasing spectral index, the predicted flux at low energies increases and hence the conversion efficiency $\xi_\mathrm{max}$ decreases to remain compatible with the upper limit on the \gr\ emission. For an injection index of $s = 2.5$ the maximum conversion fraction is below 1\%.

For an injection index $s$ below 2, the maximum energy $E_\mathrm{max}$ becomes important in the calculation of the injection rate $Q_0$ (see eq.~\ref{inject2}). For instance, for an injection index of $s=1.5$ the relation is $Q_0 \propto \sqrt{E_\mathrm{max}}$. As the maximum conversion factor $\xi_\mathrm{max}$ depends linearly on the injection rate, it also significantly rises. For $E_\mathrm{max}=10\,\mathrm{PeV}$, $\xi_\mathrm{max}$ is about 16\% for the Rosette Nebula and 23\% for Orion Nebula. However, such indices require very specific origins; they could for instance be produced by collective shock acceleration \citep{Klepach00, Bykov01}.

\subsection{Sensitivity of CTA to the selected star clusters}

Figures~\ref{fig:ModelFitsRosette} and \ref{Fig:orion_Time} show that the predicted \gr\ emission extends up to energies of several tens of TeV, which is an energy range accessible to CTA \citep{CTA1,CTA2}.

With the one-zone model presented here and assuming realistic parameters, the minimum conversion efficiency $\xi_{CTA}$ allowing a detection with CTA can be estimated. If the conversion efficiency for a star cluster is larger than $\xi_{CTA}$ it will be detected with a statistical significance of more than $5\,\sigma$ after 50\,h of observations \citep[using the sensitivity curves of ][]{2012AIPC1505769B}. These minimum efficiencies for the selected set of star clusters are summarized in Table~\ref{CTAtable}.

A detection is possible with a conversion efficiency as low as 1\% for most of the clusters. NGC\,2244 and NGC\,1976, in particular, are still ideal targets for CTA because many of the input parameters for the modelling are constricted by existing observations. In the case of non-detection, the limits on the conversion efficiency could be significantly improved.

\begin{table}[tp!]
  \caption{Minimal efficiencies for a $5\,\sigma$ detection after 50\,h observation time with the future Cherenkov Telescope Array \citep[CTA; ][]{2012AIPC1505769B} assuming realistic parameters (see section \ref{paramresum}). The star cluster RCW\,8 is not detectable.}
\centering

%

%

\begin{minipage}[t]{0.29\textwidth}
\begin{tabular}{l | c}

\hline \hline
Cluster   & $\xi_{CTA}$ (\%)\\
\hline        \hline
 NGC\,2244              &       0.72    \\

 NGC\,2467              &       4.21\\
 NGC\,6618              &       0.37\\
 RCW\,8         &       N.D.            \\
\hline    \hline
\end{tabular}
\end{minipage}
\begin{minipage}[t]{0.19\textwidth}
\begin{tabular}{l | c}
\hline \hline
Cluster   & $\xi_{CTA}$ (\%)\\
\hline        \hline
 NGC\,1976              &       1.13            \\
 NGC\,2175              &       2.21            \\
 RCW\,62                 &      0.03            \\
 NGC\,3324              &       16.1            \\
\hline        \hline
\end{tabular}
\end{minipage}

\label{CTAtable}
\end{table}

\section{Conclusions}
The \gr\ emission from young stellar clusters embedded in molecular clouds is modelled with a one-zone model, where the stellar wind luminosity is converted into relativistic protons and electrons that subsequently produce electro-magnetic radiation through synchrotron radiation, inverse Compton upscattering of low-energy photon fields, and decay of pions produced in hadronic interactions. The predictions are then compared with Fermit-LAT data and with the CTA sensitivity. The main conclusions obtained are:\ \textit{i)} Pion decay is clearly the dominating process for \gr\ production. \textit{ii)} We selected eight embedded stellar clusters based on their astrophysical properties corresponding to the presented one-zone model. No significant \gr\ emission from these star clusters was found with the \fermi in the energy range from 3 to 300\,GeV. \textit{iii)} The upper limits on the \gr\ emission from these clusters show that not more than about 10\% of the stellar wind luminosity is converted into relativistic particles. Certain clusters must have an even lower acceleration efficiency of less then 1\%. These results are reliable considering realistic parameters in the modelling. Only injection spectra harder than $s=2$ could provide acceleration efficiency higher than 10\%. However, the results presented here only consider  young stellar clusters. It cannot be excluded that the stellar wind luminosity is converted much more efficiently in older stellar clusters containing more evolved stars or supernova remnants. \textit{iv)} The CTA would be able to detect \gr\  emission from the clusters NGC\,2244, NGC\,1976, NGC\,2467, or NGC\,2175 in the case of an acceleration efficiency of close to one percent.

\bibliography{ESC.bib}
\bibliographystyle{aa}

\end{document}